\documentclass[twocolumn]{pasj02}

\jyear{2025}
\Received{2025/04/14}
\Accepted{2025/07/01}
\Published{$\langle$publication date$\rangle$}
\usepackage[switch,mathlines]{lineno}

\usepackage{url}
\usepackage{comment}
\usepackage{multirow}
\usepackage{graphicx}	
\usepackage{amsmath}	

 \def\/{\over}\def\kms{km s$^{-1}$}

 \def\be{\begin{equation}} 
 \def\ee{\end{equation}}
 \def\kms{km s$^{-1}$}    
    
\def\({\left(} \def\){\right)} \def\[{\left[} \def\]{\right]}

\def\vlsr{V_{\rm LSR}}

\def\kms{km s$^{-1}$\ }

\everymath{\displaystyle}

\begin{document}

\title{A catalog of molecular clouds {possibly associated with} Galactic infrared bubbles I. The Southern Galactic plane}

\author{
 Mikito \textsc{Kohno}\altaffilmark{1,2}\altemailmark,\orcid{0000-0003-1487-5417} \email{kohno.ncsmp@gmail.com} \email{mikito.kohno@gmail.com}
 Yoshiaki \textsc{Sofue}\altaffilmark{3},\orcid{0000-0002-4268-6499}
 Yasuo \textsc{Fukui}\altaffilmark{2,4},\orcid{0000-0002-8966-9856}
 and 
 Kengo \textsc{Tachihara}\altaffilmark{2}\orcid{0000-0002-1411-5410}
}
\altaffiltext{1}{Curatorial division, Nagoya City Science Museum, 2-17-1 Sakae, Naka-ku, Nagoya, Aichi 460-0008, Japan}
\altaffiltext{2}{Department of Physics, Graduate School of Science, Nagoya University, Furo-cho, Chikusa-ku, Nagoya, Aichi 464-8602, Japan}
\altaffiltext{3}{Institute of Astronomy, The University of Tokyo, 2-21-1 Osawa, Mitaka, Tokyo 181-0015, Japan}
\altaffiltext{4}{Faculty of Engineering, Gifu University, 1-1 Yanagido, Gifu, Gifu 501-1193, Japan}


\KeyWords{ISM: bubbles --- ISM: clouds --- ISM: molecules --- ISM: radio lines --- ISM: general}

\maketitle

\begin{abstract}
We have carried out a morphological search for molecular clouds {possibly associated with} 48 Galactic infrared bubbles with angular radii of $>\timeform{1'}$ in the southern Galactic plane of $\timeform{295D}\le l\le \timeform{350D}$ and $|b|\le \timeform{1D}$ presented by \citet{2019PASJ...71....6H}.
{116 molecular clouds
in the $(l,b,\vlsr)$ space}
are identified from the archival $^{12}$CO~$J$~=~1--0 line data obtained by the Mopra Southern Galactic plane survey,
{where $\vlsr$ is the CO-line radial velocity}. 
The kinematic distances are derived from $\vlsr$
using the most accurate rotation curve of the Milky Way. We also present measurements of velocity dispersion, size parameter, molecular mass, and virial mass of 116 molecular clouds.
This catalog is presented to investigate star formation and the origin of molecular shells/cavities possibly associated with infrared bubbles. 
 CO line intensity maps and position-velocity diagrams of the molecular clouds are available online as supplementary data.
\end{abstract}


\section{Introduction} 
Ionized hydrogen (H\emissiontype{II}) regions are formed around OB-type stars, mainly distributed on the disk of galaxies \citep{2014ApJS..212....1A,2019ApJ...871..145A}. 
Due to stellar winds and ultraviolet radiation, the OB stars ionize and destroy parent molecular clouds and form cavity-like structures around the exciting stars (\cite{1979A&A....71...59T,1979MNRAS.186...59W,2012MNRAS.427..625W,2013MNRAS.436.3430D}). 
 Cavity-like structures in the interstellar medium (ISM) are called interstellar bubbles (\cite{1975ApJ...200L.107C,1977ApJ...218..377W}).
H\emissiontype{II} regions are also known as sites of triggered star formation \citep{1998ASPC..148..150E}. They induce second-generation star formation due to shock compression by expanding bubbles \citep{2006A&A...446..171Z,2006ApJ...646..240H,2010A&A...523A...6D,2010ApJ...716.1478W,2011ApJ...733...17I,2014A&A...569A..36X,2015A&A...580A..49I,2021SciA....7.9511L}. 
The Spitzer satellite cataloged the infrared bubbles in the Milky Way from the Galactic Legacy Infrared Mid-Plane Survey Extraordinaire data (GLIMPSE: \cite{2006ApJ...649..759C,2007ApJ...670..428C}). The authors identified about 600 bubbles in the northern and southern hemispheres in the Galactic plane ($|l| \le \timeform{65D}$, $|b| \le \timeform{1D}$),
and suggest that the bubbles are H\emissiontype{II} regions with embedded OB-type stars. The Milky Way project collaborated by citizen scientists presented a catalog of 5106 infrared bubbles in the Galactic plane \citep{2012MNRAS.424.2442S,2014ApJS..214....3B,2019MNRAS.488.1141J}.
The authors revealed the position, radius, and thickness of each infrared bubble.

\citet{2016PASJ...68...37H} presented the total infrared luminosity of these infrared bubbles using AKARI all-sky survey infrared data. Hanaoka et al. (2019,2020) extended these studies and derived the shell radii and the covering fractions of infrared bubbles based on AKARI and Herschel data. More recently, \citet{2025PASJ..tmp...21N} developed a new identification method for infrared bubbles by applying deep learning to the infrared image data.
The kinematics of molecular shells and arc-like structures associated with individual infrared bubbles have been studied by CO-line analysis \citep{2010ApJ...709..791B,2012A&A...544A..39J,2015ApJ...800..101A,2015ApJ...806....7T,2016AJ....152..117Y,2016ApJ...833...85B,2018PASJ...70S..46F,2016ApJ...833...85B,2016ApJ...826...27D,2016A&A...585A..30C,2016ApJ...830...57G,2017RMxAA..53...79C,2017ApJ...851..140D,2018ApJ...866...20D,2018PASJ...70S..47O,2018PASJ...70S..51T,2019ApJ...872...49F,2019MNRAS.487.1517L,2021SciA....7.9511L,2021PASJ...73S.368T,2023MNRAS.525.4540S,2023PASJ...75..397K}.

However, systematic studies of parent molecular clouds {possibly associated with} infrared bubbles have not yet been conducted.
This paper presents the results of the identification of molecular clouds {spatially and kinematically close to} cataloged infrared bubbles in the southern Galactic plane.
It provides basic data for studies of star formation and the origin of molecular shells/cavities, {located near} the infrared bubbles in the Galactic disk.

This paper is constructed as follows: section 2 introduces the archival data; Section 3 presents methods of analysis and a catalog of molecular clouds; Section 4 presents the statistical results of molecular clouds; and Section 5 summarizes this paper.

\section{Data} 

\begin{table*}[h]
\caption{Data properties}
\begin{center}   
\begin{tabular}{cccccccccc}
\hline
\multicolumn{1}{c}{Telescope/Survey} & Line  & HPBW &Resolution  & Grid &  Velocity & r.m.s noise$^*$ & References \\
& & & &&Resolution$^\dag$ & level &\\
\hline
Mopra &{$^{12}$CO~$J$~=~1--0} & $\sim \timeform{33"}$ & \timeform{36"} & \timeform{30"} &1 km s$^{-1}$& $\sim 0.7 $ K & [1,2,3,4] \\
& && &\\
\hline
 & Band  & Detector &Resolution& Grid && & References  & \\
\hline
{Spitzer}/GLIMPSE & 8.0 $\mu$m & IRAC  &$\sim$ \timeform{2"} & \timeform{1.2"} & && [5,6,7] & \\
{Spitzer}/MIPSGAL  & 24 $\mu$m & MIPS  & \timeform{6"} &  \timeform{1.25"} & && [8,9]  &\\
\hline
\end{tabular}
\label{obs_param}
 \vspace{3pt}
\label{obs_para}\\
{\raggedright Notes. 
$^\dag$ We used the 'giant cube' binned to a velocity resolution of 1 \kms.
$^*$The r.m.s noise level is taken from the final cube data using this paper. References [1] \citet{2013PASA...30...44B}, [2] \citet{2015PASA...32...20B}, [3] \citet{2018PASA...35...29B}, [4] \citet{2023PASA...40...47C}, [5] \citet{2003PASP..115..953B}, [6] \citet{2004ApJS..154...10F}, [7] \citet{2009PASP..121..213C}, [8] \citet{2009PASP..121...76C}, [9] \citet{2004ApJS..154...25R}
 \par}
 \end{center}
\end{table*}

We used the {$^{12}$CO $J =$ 1-0} archival Galactic plane CO survey data obtained by the Mopra 22 m telescope \citep{2013PASA...30...44B,2015PASA...32...20B,2018PASA...35...29B,2023PASA...40...47C}\footnote{https://mopracosurvey.wordpress.com}. The survey area is $\timeform{250D}\le l\le \timeform{355D}$ and $|b|\le \timeform{1D}$ \citep{2023PASA...40...47C}.
The Mopra telescope is operated by the Australia Telescope National Facility (ATNF) in the Warrumbungle Mountains. 
The half-power beam width of Mopra at 115 GHZ is \timeform{33"}. The survey data are converted to the main-beam temperature ($T_{\rm MB}$) from the antenna temperature ($T_{\rm A}^*$) using the relation of $T_{\rm MB} = T_{\rm A}^*/\eta$ with the extended beam efficiency of $\eta=0.55$ \citep{2005PASA...22...62L}. Some observations were made remotely from Nagoya University in Japan using an internet connection.
The spatial and velocity resolutions of the data release 4 (DR4) are \timeform{36"} and 0.1 \kms, respectively.
We used "giant cubes", which are re-binned the Mopra archival cube to the velocity resolution of 1 \kms. 
The grid size of the cube data used in this article is $(l,b,v)=$ (30\arcsec, 30\arcsec, 1 \kms).
{The noise levels of the root mean square are $\sim 0.7$ K in $^{12}$CO $J=$1--0.}
The Mopra CO data were retrieved from the portal page of https://doi.org/10.25919/9z4p-mj92.

We also utilized the archival infrared data of the Spitzer Space telescope with the band of $8\ \mu$m \citep{2003PASP..115..953B,2009PASP..121..213C}, and $24 \ \mu$m \citep{2009PASP..121...76C}. 
The spatial resolutions of the $8\ \mu$m and $24\ \mu$m data are \timeform{2"} and \timeform{6"}, respectively.
The data properties are summarized in Table \ref{obs_para}.

\section{A catalog of molecular clouds}
\begin{figure*}[h]
 \includegraphics[width=16cm]{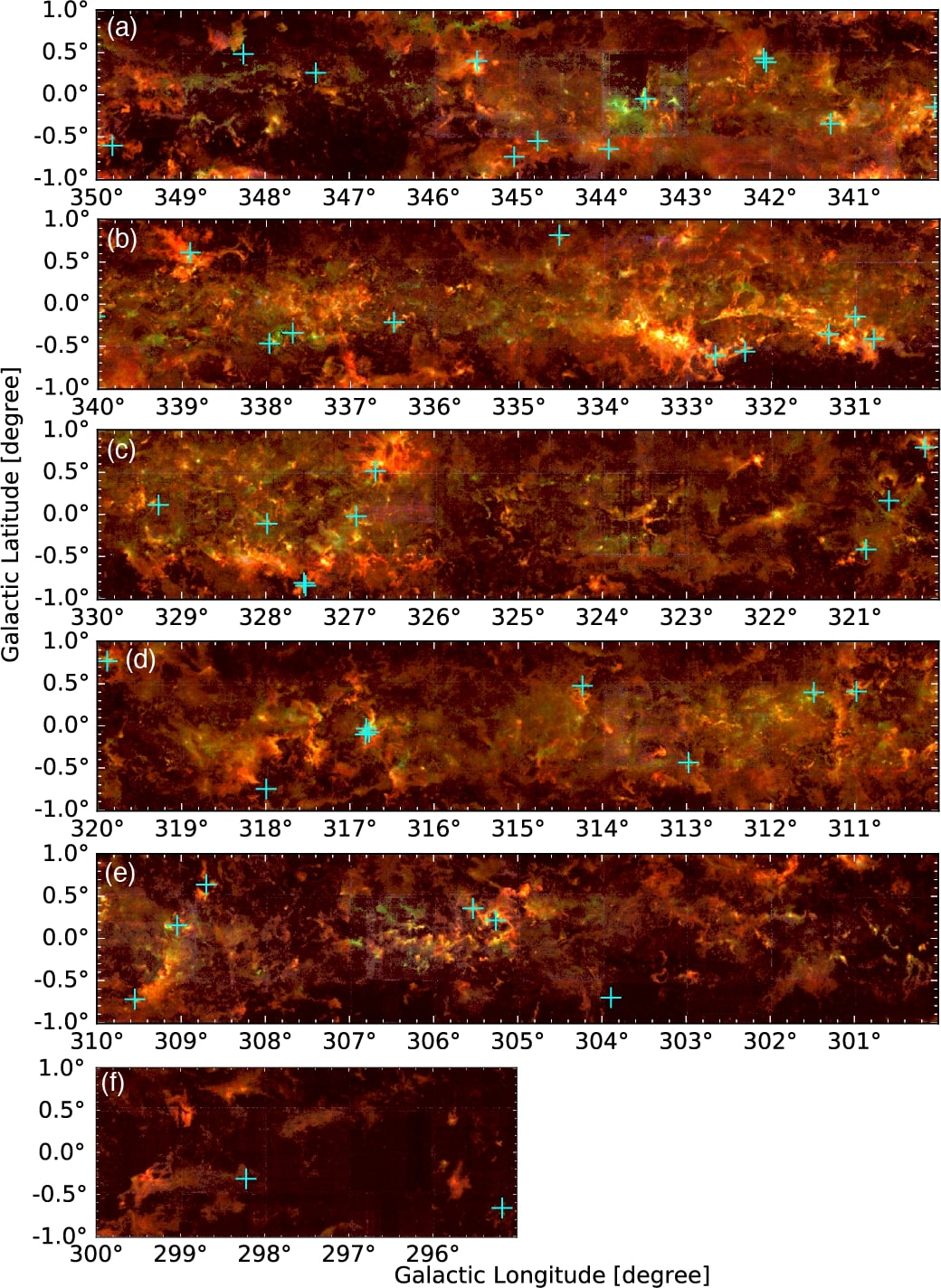}
\caption{The three-color composite image of the CO peak brightness temperatures obtained by Mopra \citep{2023PASA...40...47C}. Red, green, and blue present the $^{12}$CO, $^{13}$CO, and C$^{18}$O $J=$ 1--0 lines, respectively. Cyan crosses show the position of infrared bubbles with the angular radii of $>\timeform{1'}$ presented by \citet{2019PASJ...71....6H}. The Galactic longitude ranges are of (a) $\timeform{340D}\le l \le \timeform{350D}$, (b) $\timeform{330D}\le l \le \timeform{340D}$ (c) $\timeform{320D}\le l \le \timeform{330D}$, (d) $\timeform{310D}\le l \le \timeform{320D}$, (e) $\timeform{300D}\le l \le \timeform{310D}$, and (f) $\timeform{295D}\le l \le \timeform{300D}$ respectively. {Alt text: CO peak brightness temperature map.}}
\label{threecolor}
\end{figure*}

\subsection{Identification}
\begin{figure*}[h]
 \includegraphics[width=16.5cm]{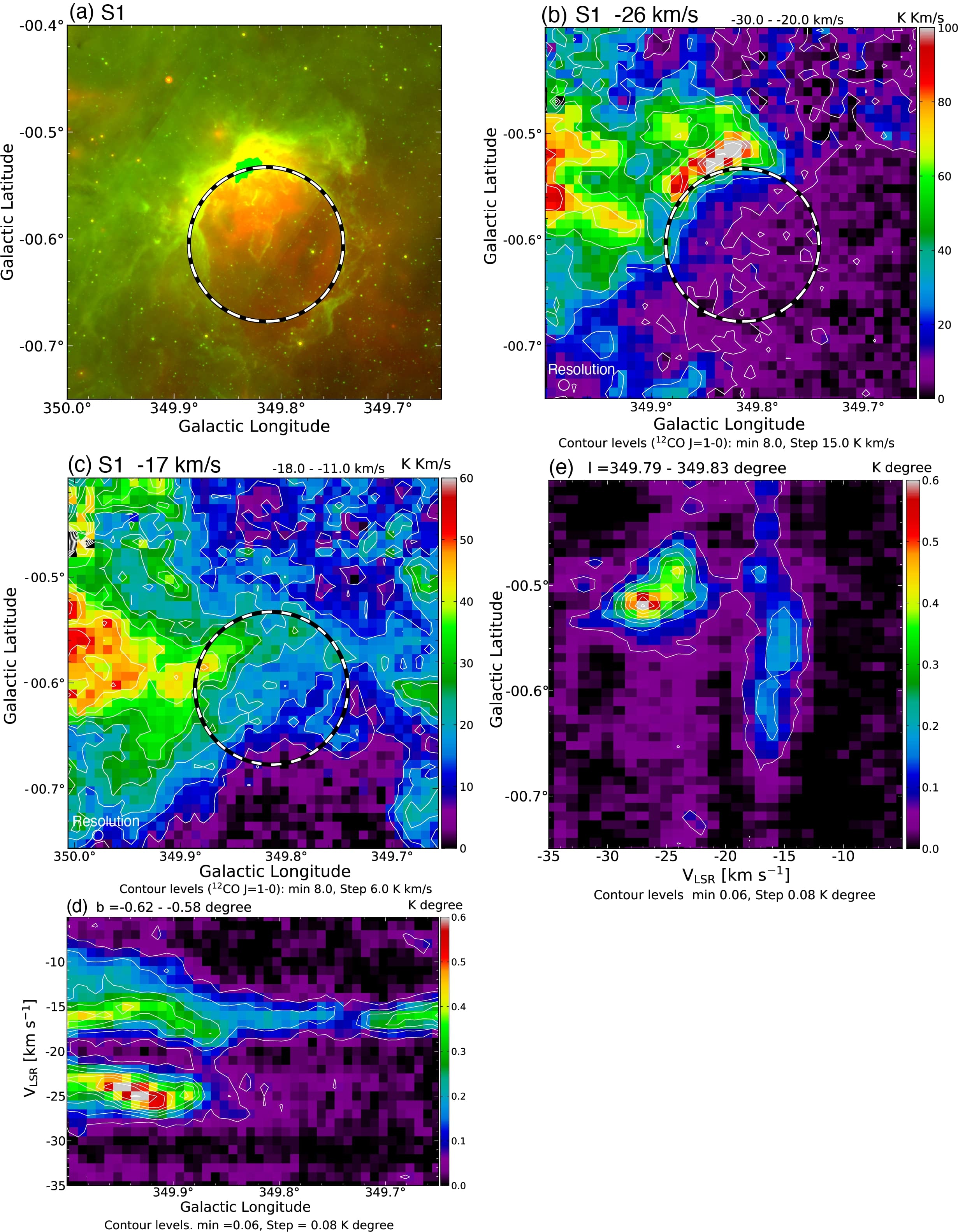}
\caption{(a) The two-color composite infrared image of the S1 bubble. The green and red show the $8\ \mu$m and $24\ \mu$m images, respectively. (b) $^{12}$CO~$J$~=~1--0 integrated intensity map. The integrated velocity range is from $-30$~\kms to $-20$~\kms. The lowest contour levels and intervals are 8~K~\kms and 15~K~\kms, respectively. (c) Same as (b), but the integrated velocity range is from $-18$~\kms to $-11$~\kms. The lowest contour levels and intervals are 8~K~\kms and 6~K~\kms, respectively. {The velocity at the top of panels (b) and (c) indicate the radial velocity of a molecular cloud at the peak position as shown in column (5) of Table \ref{tab2}.} White dashed circles indicate the boundary of the infrared bubble derived by \citet{2019PASJ...71....6H}. (d) The longitude-velocity diagram. The integrated latitude range is from $\timeform{-0.62D}$ to $\timeform{-0.58D}$. The lowest contour levels and intervals are 0.06 K~degree and 0.08 K~degree, respectively. (e) The latitude-velocity diagram. The integrated longitude range is from $\timeform{349.79D}$ to $\timeform{349.83D}$. The lowest contour levels and intervals are 0.06 K~degree and 0.08 K~degree, respectively. {Alt text: The S1 bubble, showing the infrared composite image, $^{12}$CO integrated intensity maps, and positions-velocity diagrams.}}
\label{s1}
\end{figure*}

In this paper, we present the catalog of molecular clouds {spatially and kinematically close to} infrared bubbles with the angular radii of $>\timeform{1'}$ at $\timeform{295D}\le l\le \timeform{350D}$ and $|b|\le \timeform{1D}$ based on \citet{2019PASJ...71....6H}. 
Figure \ref{threecolor} shows the three-color composite image of CO peak brightness temperatures. Red, green, and blue represent $^{12}$CO, $^{13}$CO, and C$^{18}$O~$J$~=~1--0 lines, respectively. The cross marks show the positions of infrared bubbles searched for molecular clouds.
The search for CO clouds was performed using the following procedure.
First, we display a $^{12}$CO~$J$~=~1--0 channel map and superpose a shell boundary based on infrared data by \citet{2019PASJ...71....6H}. 
Then, the CO channel is changed {near} the radial velocity of the H\emissiontype{II} regions obtained by radio recombination line {of H109$\alpha$, H110$\alpha$, and CS $J$~=~2--1 lines corresponding to each infrared bubble reported by previous studies \citep{1987A&A...171..261C,1996A&AS..115...81B,2014MNRAS.438..426H}.} 
Recent studies reported that several molecular cloud components are physically associated with infrared bubbles \citep{2009A&A...494..987P,2015ApJ...806....7T}, but their velocity {ranges} were not uniform. So, in our search, we adopted a uniform velocity range of $\sim \pm 20$ \kms {located near} each infrared bubble. 
These processes were applied to all the 48 infrared bubbles. 

Figure \ref{s1}(a) presents the two-color composite infrared image of the S1 bubble. Green and red show the $8\ \mu$m and $24\ \mu$m images obtained by the Spitzer space telescope, respectively.
These bands trace the polycyclic aromatic hydrocarbons (PAH) emission in the photo-dissociation region \citep{2003ARA&A..41..241D,2007ApJ...657..810D} and hot dust heated by OB-type stars \citep{2009PASP..121...76C}. 
Figures \ref{s1}(b) and (c) show the $^{12}$CO~$J$~=~1--0 integrated intensity maps of identified molecular clouds. The black circles present the boundary of the infrared shell defined by the AKARI $9\ \mu$m data \citep{2019PASJ...71....6H}.
{Molecular clouds either distribute along the shell edge (Figure \ref{s1}b) or exist inside the shell (Figure \ref{s1}c).
Figures \ref{s1}(d) and (e) demonstrate the longitude-velocity and latitude-velocity diagrams, respectively. 
The two velocity components connect in the velocity space at $(l,b)=(\timeform{349.87D}, \timeform{-0.50D})$ (Figures \ref{s1}d and \ref{s1}e).}
We present the Spitzer infrared image, CO spatial distributions of each radial velocity component, and position-velocity diagrams in 47 other infrared bubbles as supplementary figures E1--E43.
{In particular, the identified CO cloud close to S8 has a ring-like structure on the position-velocity diagram, which is different from other molecular clouds (supplementary figure E2d).}

\subsection{Kinematic distances, diameters, and size parameters}
\begin{figure*}[h]
 \includegraphics[width=18cm]{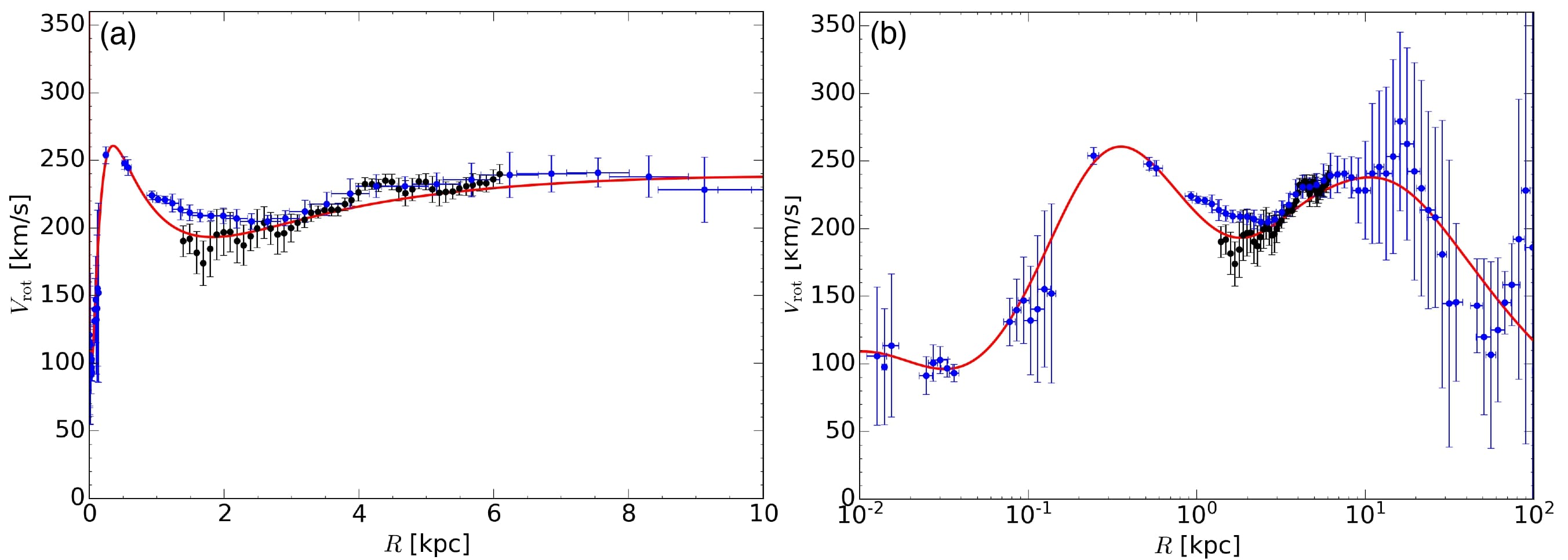}
\caption{(a) Most accurate rotation curve of the Milky Way from the Galactic center to 10 kpc \citep{2023Ap&SS.368...74S}. (b) The logarithmic rotation curve from $10^{-3}$ to $10^2$ kpc. Black dots were obtained by the FUGIN CO survey data \citep{2021PASJ...73L..19S}. Blue dots were taken from a compilation for the entire Galaxy from the nucleus to the halo \citep{2013PASJ...65..118S}. The error bar of the rotation velocity is estimated from the intrinsic fluctuation and errors of the rotation curve. The red lines show the model curve using the kinematic distance determination in this paper. {Alt text: Rotation curve of the Milky Way.}}
\label{rotcurve}
\end{figure*}

\begin{figure*}[h]
 \includegraphics[width=18cm]{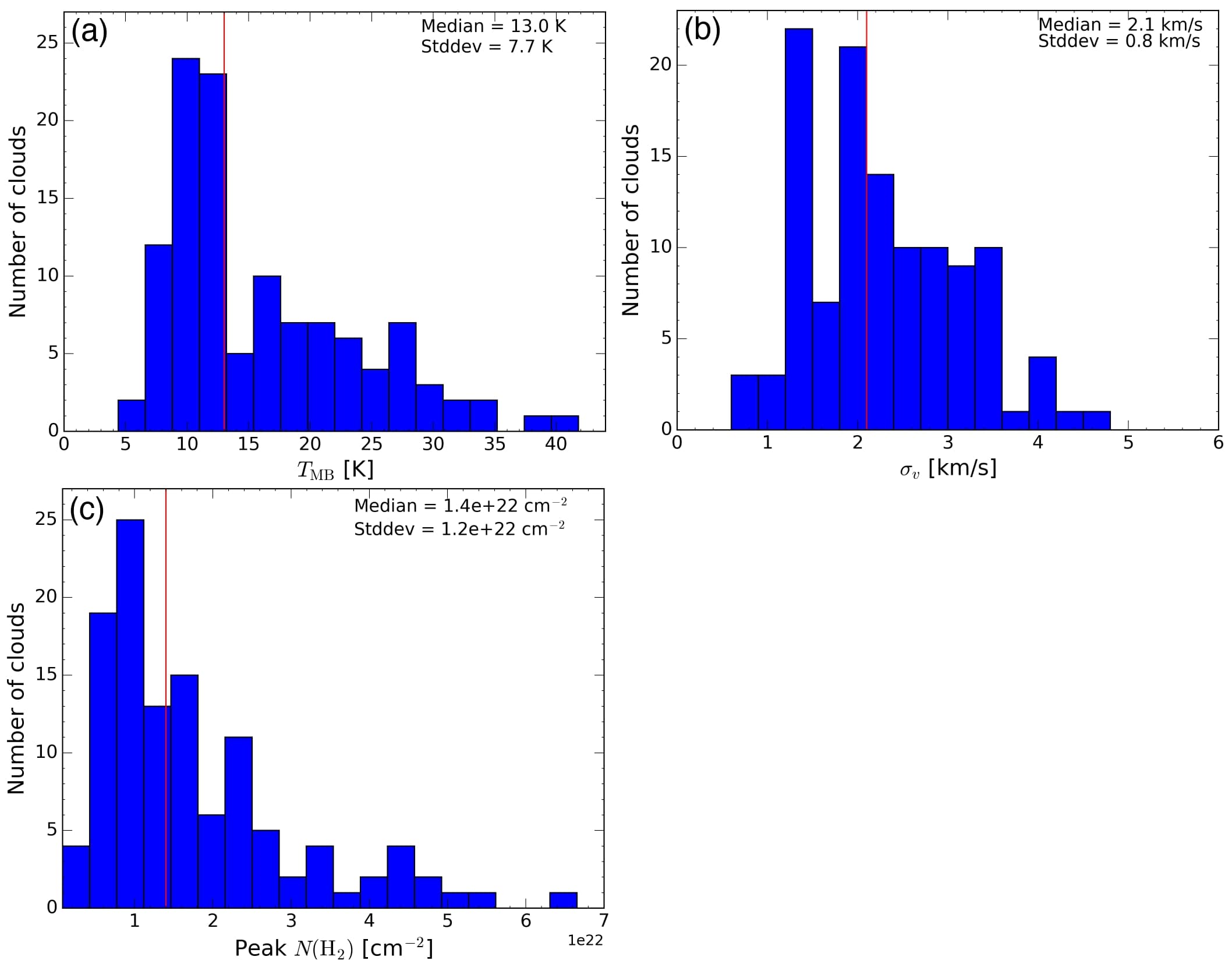}
\caption{The histograms of peak brightness temperature in panel (a), velocity dispersion ($\sigma_v$) in panel (b), and peak H$_2$ column density in panel (c). The vertical red lines indicate the median values. "Stddev" means the standard deviation of each panel. {Alt text: Histograms of peak brightness temperature, velocity dispersion ($\sigma_v$), and peak H$_2$ column density. }} 
\label{hist}
\end{figure*}

{If we assume the circular rotation of the molecular gas in the Galactic disk, the radial velocity $V_{\rm LSR}$ is given by
\begin{equation}
V_{\rm LSR} = \left(\frac{V(R)}{R} - \frac{V_{0}}{R_{0}} \right)R_{0}\sin{l},
\end{equation}
where $V(R)$ is the circular rotation velocity as a function of the Galactocentric distance $R$, $V_0$ is the circular velocity of the local standard of rest (LSR) at the Sun, $R_{0}$ is the distance from the Galactic center to the Sun, and $l$ is the Galactic longitude. We derived $V(R)$ from the rotation curve of the Milky Way. In this paper, we adopted the high-accuracy rotation curve of the Milky Way because it is the sensitive parameter to derive the kinematic distance from the radial velocity \citep{2023Ap&SS.368...74S}. The Galactic constants are adopted to $R_{0}= 7.92$ km s$^{-1}$ and $V_{0}= 227$ km s$^{-1}$ obtained by the Very Long Baseline Interferometry (VLBI) astrometry observations \citep{2020PASJ...72...50V}. 
 Figures \ref{rotcurve}(a) and (b) show the rotation curve from the Galactic center to 10 kpc and from $10^{-3}$ to $10^{2}$ kpc in the logarithmic scale based on \citet{2023Ap&SS.368...74S}. 
The red line presents a rotation curve using this paper based on the fitting result of recent observational data \citep{2013PASJ...65..118S,2021PASJ...73L..19S}.
The kinematic distance $d$ is given by
\begin{equation}
d = R_{0}\cos{l} \pm \sqrt{R^{2} - R_{0}^{2}\sin^2{l}}.
\end{equation}
}
We adopted molecular clouds in this catalog to satisfy the relation of $R^{2} - R_{0}^{2}\sin^2{l}>0$. For a given radial velocity, the kinematic distance has two solutions at the near and far sides of the inner Galaxy. 
The errors of kinematic distances $\delta d$ are estimated from equation (2) with the uncertainty of the radial velocity of $\delta v_{\rm LSR} \pm 5$ \kms caused by interstellar turbulence. 
{Recently, \citet{2023PASJ...75..279F} developed the machine-learning-based distance estimation methods to solve the near-far kinematic distance ambiguity. This method will be helpful to revise the uncertainties of our study for the future.}

The linear diameter of the infrared bubble $D$ is expressed as
\begin{equation}
D=d\ {\rm tan}(2r),
\end{equation}
where $r$ is the radii of the infrared bubble taken from the AKARI $9\ \mu$m data \citep{2019PASJ...71....6H}. 
We note that the kinematic distance of each bubble and the infrared image determine the linear diameter $D$.

The centroid position along
each axis of the Galactic longitude ($l$),  Galactic latitude ($b$), and velocity axis ($v$) is calculated as
\begin{equation}
\langle l \rangle = {\displaystyle\sum_{i} T_i l_i \over \displaystyle\sum_{i} T_i}, 
\label{mom1}
\end{equation}
\begin{equation}
\langle b \rangle = {\displaystyle\sum_{i} T_i b_i \over \displaystyle\sum_{i} T_i} 
\label{mom1}
\end{equation}
, and
\begin{equation}
\langle v \rangle = {\displaystyle\sum_{i} T_i v_i \over \displaystyle\sum_{i} T_i}.
\label{mom1}
\end{equation}
The cube data has longitude $l_i$, latitude $b_i$, velocity $v_i$, and brightness temperature $T_i$ in the $i$th pixel. $\displaystyle\sum_{i}\ T_i$ is the $^{12}$CO $J=$1--0 total brightness temperature within the cloud.
We calculated within the range above $T_{\rm min}=3.5$ K,  which is the $5\sigma$ noise level with the $^{12}$CO $J$~=~1--0 data. 
The intensity-weighted standard deviation along ($l,b,v$) is defined as
\begin{equation}
\sigma_l = \sqrt{{\displaystyle\sum_{i} T_i (l_i - \langle l\rangle)^2 \over \displaystyle\sum_{i} T_i}} = \sqrt{\langle l^2 \rangle - \langle l \rangle^2},
\label{mom2}
\end{equation}

\begin{equation}
\sigma_b = \sqrt{{\displaystyle\sum_{i} T_i (b_i - \langle b\rangle)^2 \over \displaystyle\sum_{i} T_i}} = \sqrt{\langle b^2 \rangle - \langle b \rangle^2}
\label{mom2}
\end{equation}
, and
\begin{equation}
\sigma_v = \sqrt{{\displaystyle\sum_{i} T_i (v_i - \langle v\rangle)^2 \over \displaystyle\sum_{i} T_i}} = \sqrt{\langle v^2 \rangle - \langle v \rangle^2}.
\label{mom2}
\end{equation}

The integration ranges of $l$, $b$, and $v$ correspond to the longitude, latitude, and integrated velocity range in each $^{12}$CO~$J=$~1--0 integrated intensity map in the figure.
The relation between $\sigma_v$ and full-width half maximum (FWHM) line-width $\Delta V$ is $\Delta V=\sqrt{8 \ln 2}\ \sigma_v$ \citep{2006PASP..118..590R}.

The size parameter S of the molecular clouds is defined by
\begin{equation}
S = d \tan (\sqrt{\sigma_l\ \sigma_b}).
\label{size}
\end{equation}
The effective radius of the molecular cloud ($R_{\rm e}$) is related to the size parameter S by $R_{\rm e} \sim 3.4S/\sqrt{\pi}\simeq 1.92 S$ \citep{1987ApJ...319..730S}.
We derived the size parameters S from the near and far solutions of the kinematic distance. 
We note that the radius $r$ of infrared shells and the radius of molecular clouds often do not coincide because molecular clouds spatially extend more than the 9 $\mu$m ring defined by the AKARI data.


\subsection{The column density, molecular mass, and virial mass}
The H$_2$ column density assuming the CO-to-H$_2$ conversion factor ($X_{\rm CO}$) is given by
\begin{equation}
N({\rm H_2}) = X_{\rm CO}\ I_{\rm ^{12}CO}\ {\rm [cm^{-2}]},
\label{XCO}
\end{equation}
where $I_{\rm ^{12}CO}$ is the integrated intensity of $^{12}$CO $J =$ 1--0. We used $X_{\rm CO} = 2 \times 10^{20}\ {\rm [cm^{-2}\ (K\ km\ s^{-1})^{-1}]}$ as the standard value of the Galactic disk in the Milky Way \citep{2013ARA&A..51..207B,2020MNRAS.497.1851S,2024MNRAS.527.9290K}.

The CO luminosity and total molecular cloud mass are given by
\begin{equation}
L_{\rm CO} = d^2 \Omega I_{\rm ^{12}CO} \ {\rm [K\ km\ s^{-1}\ pc^2]},
\label{mass}
\end{equation}
and
\begin{equation}
M_{\rm cloud} = \mu_{\rm H_2} m_{\rm H} d^2 \Omega \sum_{i} N_i(\rm{H_2})\ {\rm [M_{\odot}]},
\label{mass}
\end{equation}
where $d$ is the kinematic distance, $\Omega$ is a solid angle of a molecular cloud, $\mu_{\rm H_2}$ is the mean molecular weight per hydrogen molecule of 2.8, $m_{\rm H}$ is proton mass of $1.67 \times 10^{-24}$ g, and $N_i(\rm{H_2})$ is the H$_2$ column density in the $i$th pixel, respectively.
{If a kinematic distance $d$ has the $\pm 20\%$ uncertainties, the CO luminosity and total molecular cloud mass have about $\pm 30\%$ uncertainties by the error propagation.}
The virial mass is expressed as 
\begin{equation}
M_{\rm vir} = 3 f_p {S \sigma_v^2 \over G}\ {\rm [M_{\odot}]},
\label{virial}
\end{equation}
where $f_p=2.9$ is a projection factor assuming a cloud  density profile of $\rho(r) \propto r^{-1}$. and
$G=6.67 \times 10^{-8}$ cm$^3$ g$^{-1}$ s$^{-2}$ 
is the Newtonian constant of gravitation \citep{1987ApJ...319..730S}.
We estimate the molecular and virial masses in the cases of near- and far-kinematic distances.
Tables \ref{tab2} and \ref{tab3} summarize our results of the physical parameters.

\section{Properties of molecular clouds}
{We point out that this catalog presents molecular cloud "candidates" physically associated with infrared bubbles. Thus it does not exclusively contain clouds physically interacting with infrared bubbles.
The analysis of the CO intensity ratio between different rotational transition levels might be useful in investigating molecular clouds interacting with infrared bubbles actually (e.g.,\cite{2015ApJ...806....7T,2021PASJ...73S.338K}). We will argue this analysis in the forthcoming paper.}
Figures \ref{hist} (a), (b) and (c) show histograms of the $^{12}$CO peak brightness temperature, velocity dispersion, and peak H$_2$ column density.
The median peak bright temperature of the molecular clouds is 13 K, the median velocity dispersion is 2.1 \kms, and the median peak column densities are $1.4 \times 10^{22}$ cm$^{-2}$.

Figures \ref{mass} (a), (b) and (c) show the histograms of the size parameter, the molecular mass, and the virial mass.
The blue and red histograms adopt the near and far solutions of the kinematic distance. 
The median cloud size is 4.1 pc and 13.9 pc for blue and red histograms, respectively.
The median cloud mass is $2.0 \times 10^4\ M_{\odot}$ and $2.0 \times 10^5\ M_{\odot}$ for applying the near and far solutions of the kinematic distance, respectively.
The median virial mass is $3.9 \times 10^4\ M_{\odot}$ and $1.2 \times 10^5\ M_{\odot}$ for the near and far kinematic distance, respectively.
In the case of both near and far solutions of the kinematic distances, the median values of the cloud and virial masses coincide with a factor of 2.

\begin{figure}[h]
 \includegraphics[width=8cm]{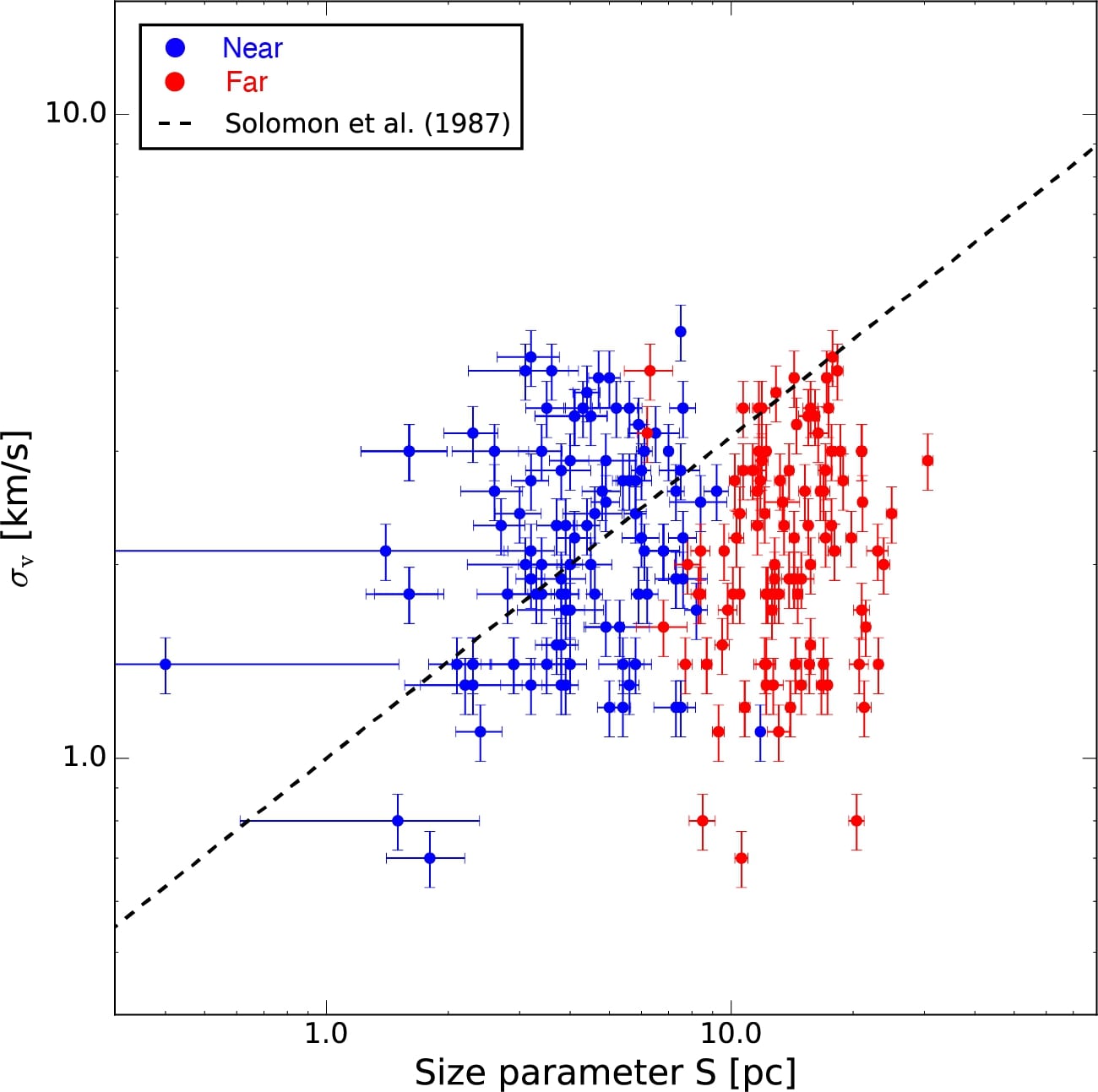}
\caption{Scatter plot between size parameter and velocity dispersion for {identified molecular clouds}. {The errors of $\sigma_v$ are assumed to the typical value as $\sim 10\% $ of a molecular cloud \citep{2006PASP..118..590R}. The errors of size parameters are calculated from the uncertainty of each kinematic distance.} Black dotted line indicates the scaling relation of $\sigma_v= S^{0.5}$ by \citet{1987ApJ...319..730S}. {Alt text: Scatter plot between size parameter and velocity dispersion.} }
\label{sizelinewidth}
\end{figure}

Figure \ref{sizelinewidth} presents the scatter plots between the molecular cloud size parameter ($S$) and the velocity dispersion ($\sigma_v$) {of our catalog}. The blue and red points correspond to the results adopting the near and far solutions of the kinematic distance. We can find almost no correlation between size and velocity dispersion, which differs from that found for Galactic molecular clouds in Virial equilibrium \citep{1981MNRAS.194..809L,2004ApJ...615L..45H}. 
This result might be caused by turbulent motions in the molecular cloud enhanced by radiative feedback from the exciting massive stars \citep{2010A&A...523A...6D} or cloud collision events around the bubble \citep{2021PASJ...73S...1F}.
{We also compared with the results of the relation of $S=\sigma^{0.5}$ by \citet{1987ApJ...319..730S} as shown in dotted line in Figure \ref{sizelinewidth}. 
The data points on the far side almost do not fit with the scaling relation of \citet{1987ApJ...319..730S}.
We suggest that this result shows the near kinematic distances are adoptable in the most infrared bubbles, same as pointed by \citet{2006ApJ...649..759C}.}

\begin{figure}[h]
 \includegraphics[width=8cm]{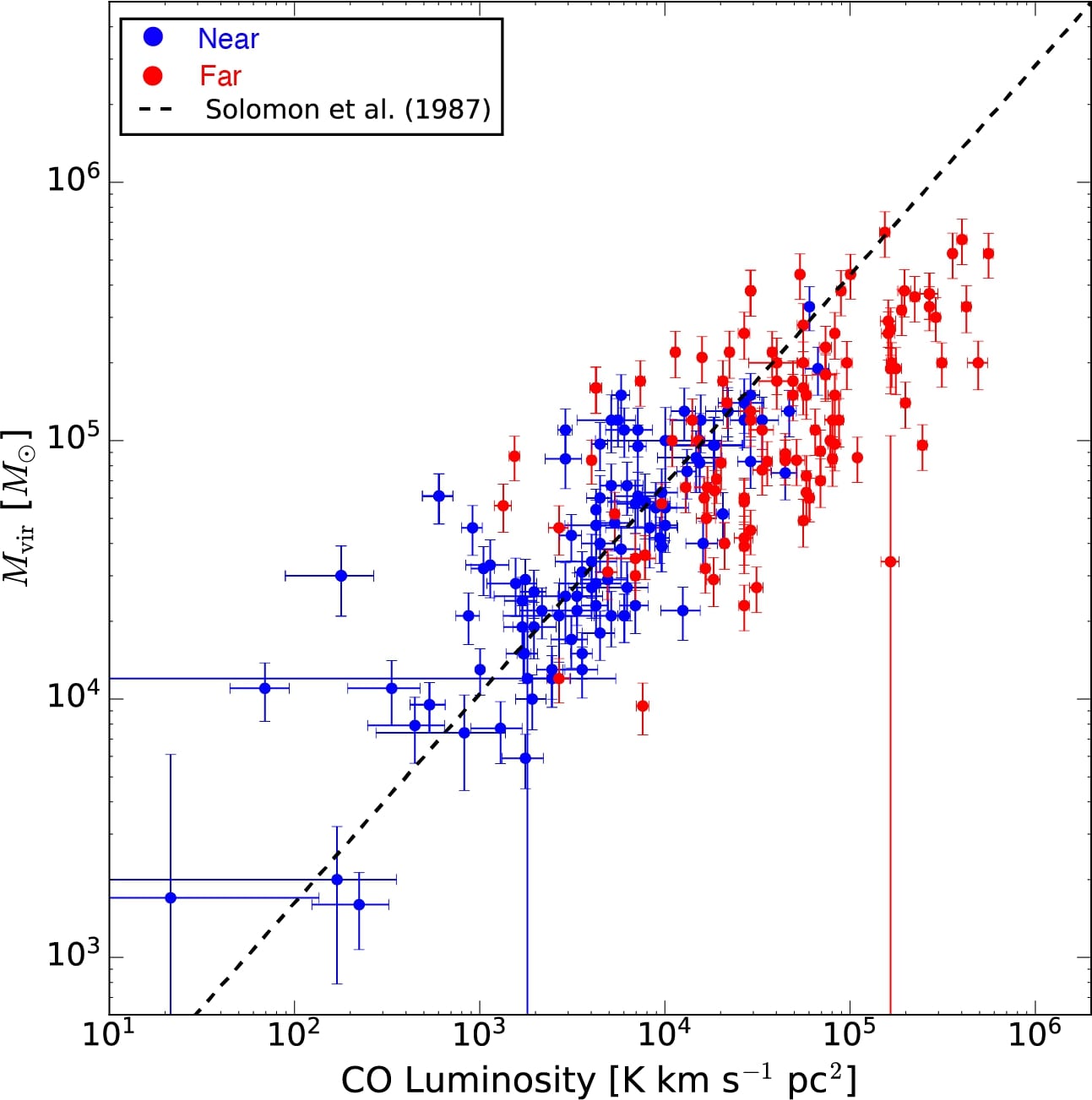}
\caption{Scatter plot between CO luminosity and virial mass. {The uncertainties of CO luminosity and virial mass are estimated from the error propagation of $\sigma_v$ and $\delta d$.} 
The black dotted line indicates the scaling relation of $M_{\rm vir}=39L_{\rm CO}^{0.81}\ [M_{\odot}]$ by \citet{1987ApJ...319..730S}. {Alt text: Scatter plot between CO luminosity and virial mass.}}
\label{luminositymass}
\end{figure}

Figure \ref{luminositymass} shows the scatter plot between CO luminosity and virial mass. 
There is a clear correlation between them. 
The ratio between CO luminosity and virial mass corresponds to the CO-to-H$_2$ conversion factor if we assume virial equilibrium \citep{2010ARA&A..48..547F,2013ARA&A..51..207B}. 
{The deviations between the data points and the relation of \citet{1987ApJ...319..730S}} could be caused by the variability of the CO-to-H$_2$ conversion factor of each molecular cloud indicated by recent studies \citep{2020MNRAS.497.1851S,2024MNRAS.527.9290K}.

\begin{figure*}[h]
 \includegraphics[width=18cm]{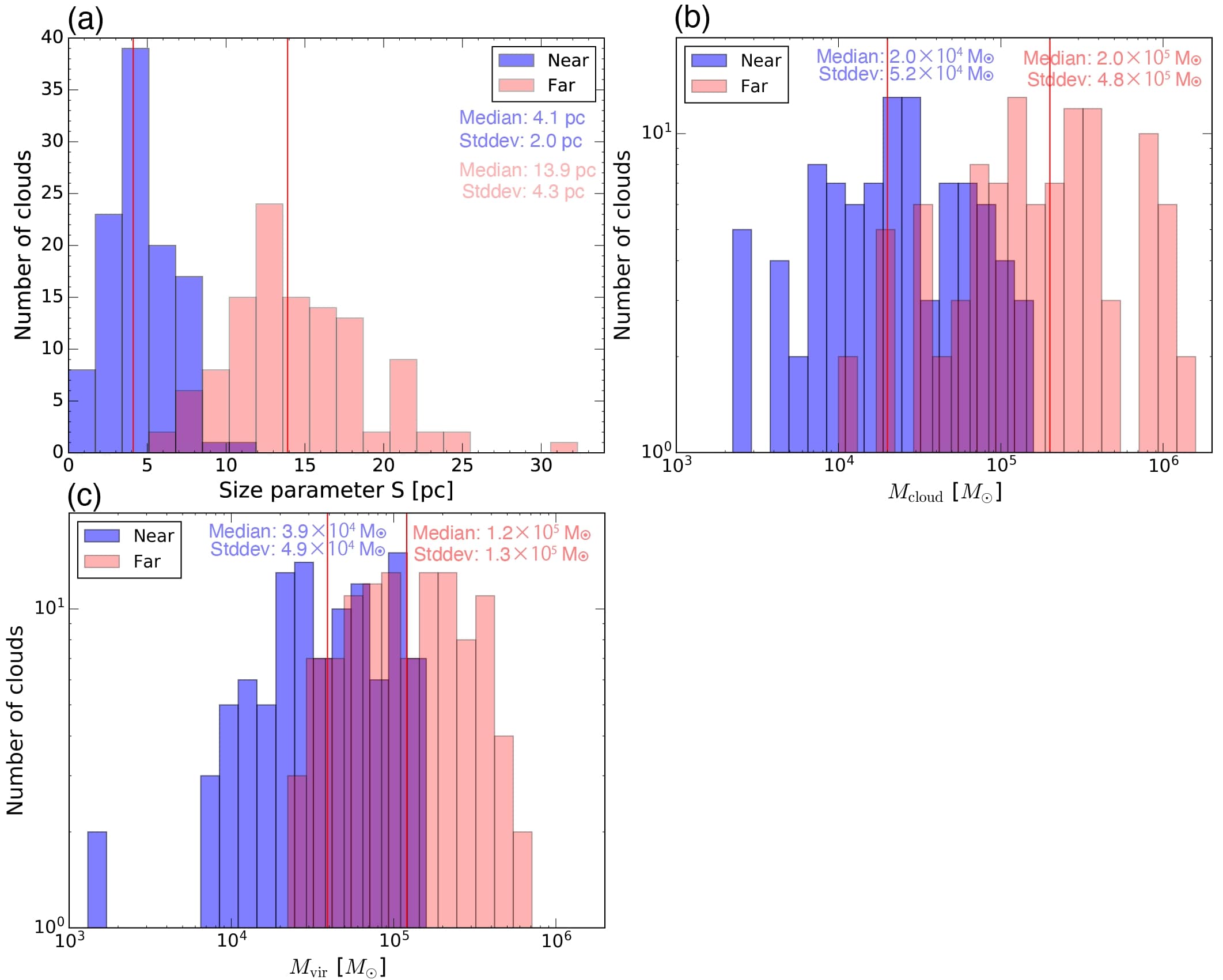}
\caption{The histograms of size parameter in Panel (a), molecular mass in Panel (b), and virial mass in Panel (c) identified molecular clouds. Blue and red show the results adopted as the near and far solution of the kinematic distance. The vertical red lines indicate the median values. {Alt text: Histograms of size parameter, molecular mass, and virial mass. }}
\label{mass}
\end{figure*}

\begin{figure}[h]
\begin{center} 
  \includegraphics[width=8cm]{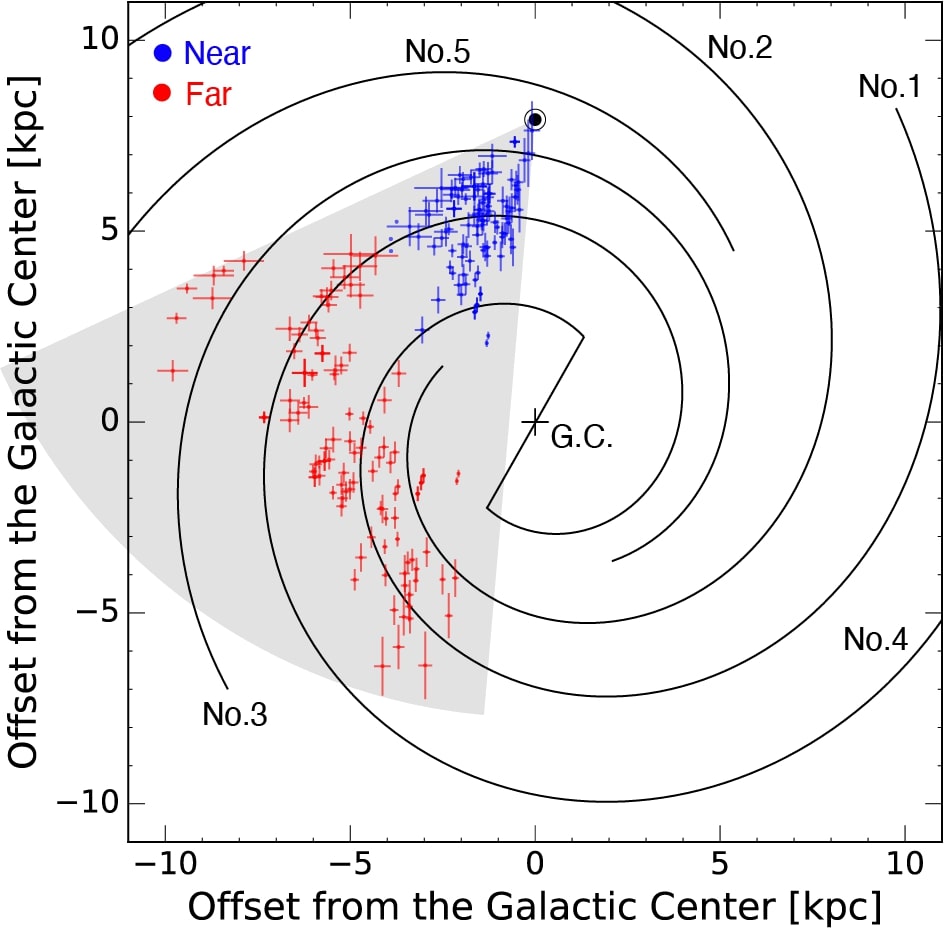}
\end{center}
\caption{Face-on distribution of CO-line clouds possibly associated with infrared bubbles in the Galactic plane. The Galactic logarithmic spiral arms are adopted by \citet{2016PASJ...68....5N}. {The errors of each molecular cloud are obtained by $\delta d$.} No.1 is the Norma--Cygnus arm, No.2 is the Perseus arm, No.3 is the Sagittarius--Carina arm, No.4 is the Scutum--Crux arm, and No.5 is the Orion (Local) arm. The grey area shows the Galactic longitude range analyzed in this paper. The $+$ and $\odot$ marks show the position of the Galactic center and Sun. {Alt text: Face-on distribution of identified molecular clouds.}}
\label{faceondistribution}
\end{figure}

Figure \ref{faceondistribution} shows the face-on distribution of molecular clouds based on the kinematic distance derived in this paper. 
The spiral arms have adopted the model by \citet{2016PASJ...68....5N}.
{Molecular clouds are distributed not only on the spiral arm but also in the inter-arm region both of the near and far solution.}

\begin{figure}[h]
 \includegraphics[width=8cm]{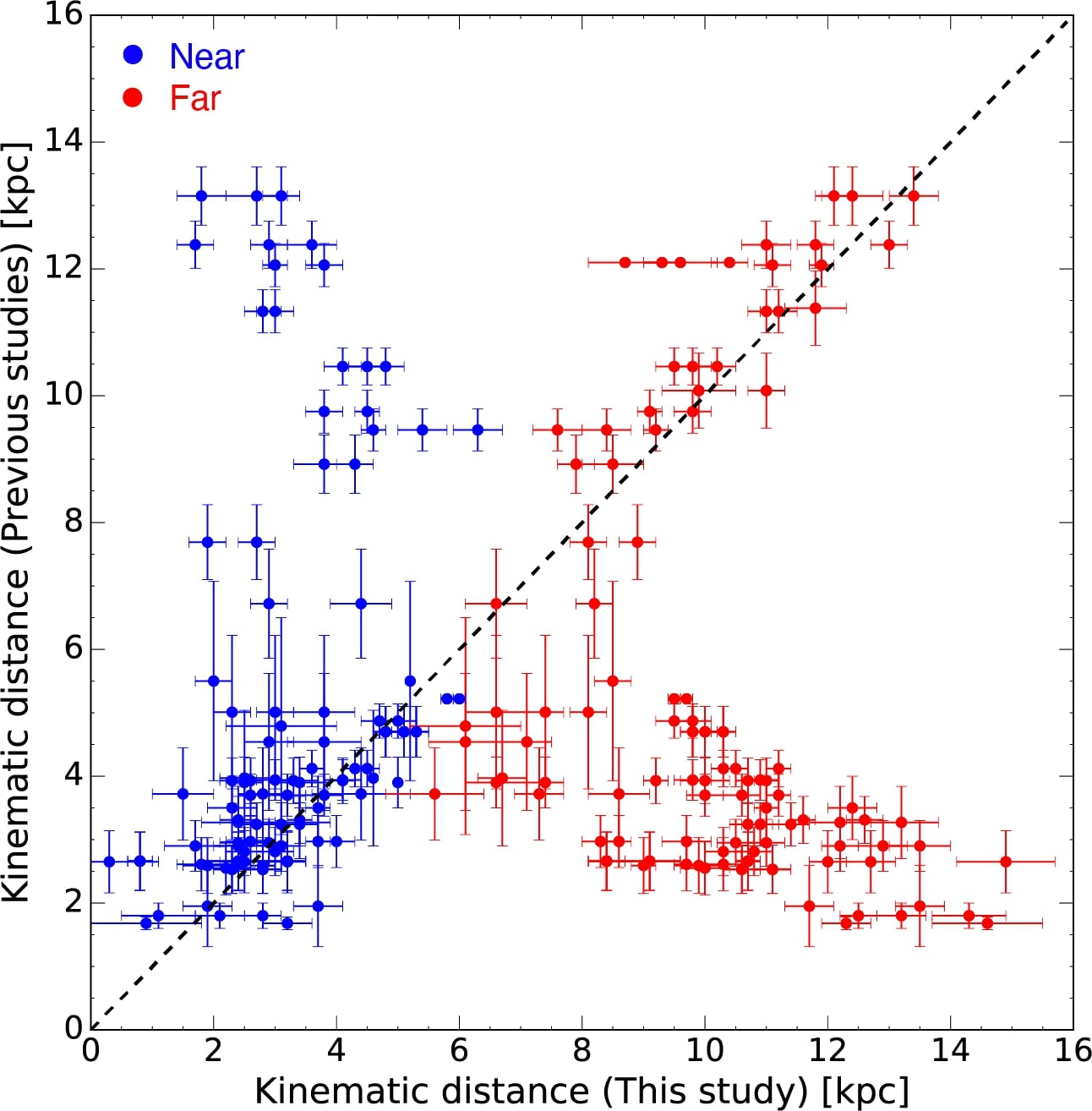}
\caption{{Comparison of kinematic distances obtained in this paper and previous studies \citep{2014MNRAS.438..426H,1987A&A...171..261C,2018A&A...617A..67S,2019ApJ...870...32K,2023A&A...679L...5B}. Blue and red circles indicate the near and far solutions of the kinematic distance in this paper.  The errors of each data point are obtained by $\delta d$ and previous studies. The black dotted line shows the relation of equal distances. Alt text: Comparison of kinematic distances obtained in this paper and previous studies.} }
\label{distance}
\end{figure}

\section{{Comparison of kinematic distances with the previous results}}
{In order to investigate the near-far ambiguity of kinematic distances obtained in this paper, we made the correlation plot of distances between this paper and previous studies.
Figure \ref{distance} shows the distance correlation plot between this study and previous works in \citet{2014MNRAS.438..426H}, \citet{1987A&A...171..261C},\citet{2018A&A...617A..67S}, \citet{2019ApJ...870...32K}, and \citet{2023A&A...679L...5B}.
Most of the molecular clouds identified in this paper coincide with the near distances, while molecular clouds possibly associated with the eleven bubbles of S17, S27, S36, S44, S54, S64, S70, S92, S163, S181, and S186 correspond to the far distances.
The distances of S7 ($=$RCW 120) and S145 ($=$ RCW 79) are derived to $1.68^{+0.13}_{-0.11}$ kpc and $3.9\pm 0.4$ kpc from non-kinematic methods
by the annual parallax measurements from Gaia satellite data \citep{2019ApJ...870...32K,2023A&A...679L...5B}, and consistent with $0.9 \pm 0.9$ kpc and $3.4 \pm 0.5$ kpc, respectively, obtained in this paper within errors. 
Thus, the near kinematic distance is reasonable in the case of S7 and S145. The annual parallax measurements of fixed stars or maser sources physically associated with infrared bubbles in the southern sky might be essential to solve the near-far ambiguity of kinematic distances, and are a potentially important topic for future studies.}
\section{Summary}
We present a molecular cloud catalog {spatially and kinematically close to} 48 Galactic round-shaped infrared bubbles {in the position-position-velocity space} at $\timeform{295D}\le l\le \timeform{350D}$ and $|b|\le \timeform{1D}$ using the archival $^{12}$CO~$J$~=~1--0 line data obtained by the Mopra Southern Galactic plane survey. We identified 116 molecular cloud components possibly associated with infrared bubbles.
The kinematic distances are calculated from the CO radial velocities using the highest-accuracy rotation curve of the Milky Way. 
We also present measurements of velocity dispersion, size parameter, molecular mass, and virial mass of 116 molecular clouds.
The CO maps and position-velocity diagrams are presented in the supplementary online figures.

\begin{table*}
\tbl{Radial velocity, peak brightness temperature,  intensity-weighted standard deviation, and peak H$_2$ column density of molecular clouds toward the infrared bubbles in the Southern Galactic plane.}{
  \centering
  \scalebox{0.95}[0.95]{
  \begin{tabular}{cccccccccccccc}
    \hline
Name & $l$ & $b$ & $V_{\rm LSR}^{\rm H\emissiontype{II}}$ & $V_{\rm LSR}^{\rm CO}$ &$T_{\rm peak}$ &$\sigma_l$&$\sigma_b$ &$\sigma_v$ & $N({\rm H2})_{\rm max}$ &  Other & References.\\
     &  [deg] & [deg]&[km s$^{-1}$] &[km s$^{-1}$]&[K]& [deg]& [deg] &[km s$^{-1}$] & [cm$^{-2}$] & Name&\\
(1) &(2) & (3) & (4) & (5) & (6) & (7) & (8) & (9) & (10)& (11)& (12)\\   
    \hline 
S1 	& 349.814 &$-0.605$& $-25$ &$-26$ & 22& 0.06 & 0.07 & 1.9 &$2.4 \times 10^{22}$&&1\\
&&&&$-17$ &13& 0.10 & 0.08 &1.7 &$1.1 \times 10^{22}$ &&\\
S7 	&348.259&${+0.483}$& $-9$ &$-7$ & 21& 0.10 & 0.08 &2.1&$2.7 \times 10^{22}$& RCW 120&2,3,4,5 \\
&&&& $-28$& 15& 0.05 & 0.04 &1.1&$9.8 \times 10^{21}$&&&\\
S8 	&347.397 &${+0.261}$&$-97$ &$-90$& 10& 0.06 &  0.06 &2.1&$1.1 \times 10^{22}$ &&1\\
&&&& $-83$&9.9& 0.08 & 0.06  &3.0 &$1.6 \times 10^{22}$ &&&\\
S11	&345.482 &${+0.400}$&$-16$ &$-18$& 38& 0.08 & 0.08 & 3.0&$6.4 \times 10^{22}$ &&1\\
& && &$-40$  & 11& 0.08 & 0.07  & 2.6 &$1.5 \times 10^{22}$ &&&\\
S13	&345.041 &$-0.737$& $-27$&$-21$ & 19& 0.14 & 0.13  & 2.9 &$2.3 \times 10^{22}$ &&6\\
&&&& $-10$& 9& 0.12 & 0.06 & 0.8&$4.0 \times 10^{21}$&&&\\
&&&& $-30$& 9& 0.10 & 0.10 & 1.6 &$8.3 \times 10^{21}$&&&\\
S14	& 344.761	&$-0.551$& $-28$ &$-33$& 11& 0.11 & 0.10  &1.4 &$8.8 \times 10^{21}$ &&7\\
&&&& $-24$& 15& 0.11 & 0.11 & 2.4&$1.4 \times 10^{22}$&&&\\
&&&& $-17$& 13& 0.10 & 0.10 & 2.0&$1.2 \times 10^{22}$&&&\\
S15	& 343.916	&$-0.647$&$-27$ &$-27$& 17& 0.09 & 0.08  & 4.0 &$3.2 \times 10^{22}$ &&1\\
&&&& $-4$& 10& 0.07 & 0.09  & 1.4&$5.8 \times 10^{21}$ &&&\\
&&&& $-36$& 11& 0.08 &  0.04 & 1.3&$7.9 \times 10^{21}$ &&&\\
S17	&	343.48	&$-0.048$&$-29$&$-30$ & 19& 0.08 & 0.08  & 2.8&$2.2 \times 10^{22}$  &&1\\
&&&& $-36$&9& 0.08 & 0.07 & 1.4&$7.7 \times 10^{21}$ &&&\\
&&&& $-19$&27& 0.10 & 0.05 & 1.3&$1.9 \times 10^{22}$ &&&\\
S18	& 342.075& ${+0.433}$& $-70$ &$-71$&13& 0.08 & 0.06 &1.8&$1.1 \times 10^{22}$&&8\\
&&&& $-79$&10& 0.09 & 0.08 &1.9&$1.0 \times 10^{22}$&&&\\
&&&& $-84$&10& 0.09 & 0.08 &1.2&$6.2 \times 10^{21}$&&&\\
S20	&	342.045	& ${+0.386}$&$-70$&$-71$&13& 0.08 & 0.06 &1.8&$1.1 \times 10^{22}$&&8\\
&&&& $-79$&10& 0.09 & 0.08 &1.9&$1.0 \times 10^{22}$&&&\\
&&&& $-84$&10& 0.09 & 0.08 &1.2&$6.2 \times 10^{21}$&&&\\
S23	&	341.28	&$-0.346$& $-42$ &$-45$ &20& 0.09 & 0.09 &3.9&$3.2 \times 10^{22}$&&1\\
&&&& $-30$&16& 0.08 & 0.08 &3.0&$1.7 \times 10^{22}$&&&\\
S27	&	340.04	&$-0.147$& $-48$ &$-53$ &23& 0.06 & 0.07 &3.7&$4.6 \times 10^{22}$&&1\\
&&&& $-39$&9& 0.06 & 0.09 &2.3&$1.1 \times 10^{22}$&&&\\
S29	&	338.906	& ${+0.612}$& $-63$ &$-66$ & 30& 0.08 & 0.08  &3.3&$5.3 \times 10^{22}$&&1\\
&&&& $-71$&10& 0.07 & 0.05 &1.8&$1.4 \times 10^{22}$&&&\\
&&&& $-52$&13& 0.10 & 0.08 &1.3&$8.8 \times 10^{21}$&&&\\
S36	&	337.965	&$-0.469$ & $-40$& $-41$ & 20& 0.09 & 0.08 & 3.5&$2.0 \times 10^{22}$&&1,9\\
&&&& $-23$& 7& 0.05 & 0.06 &1.8&$3.9 \times 10^{21}$&&\\
&&&& $-53$& 9& 0.08 & 0.04 &3.5&$1.5\times 10^{22}$&&\\
S37	&	337.688	&$-0.343$& $-48$& $-30$&17& 0.08 & 0.08 &4.2&$1.8 \times 10^{22}$&&10\\
&&& &$-54$&12& 0.09 & 0.06 &3.9&$2.3 \times 10^{22}$&&\\
S41	&	336.484	&$-0.216$&$-85$&$-84$&26& 0.06 & 0.06 &2.7&$2.7 \times 10^{22}$&&1\\
&&&& $-79$&25& 0.08 & 0.06 &2.4&$2.3\times 10^{22}$&&\\
S44	&	334.52	&${+0.818}$& $-77$ &$-80$&16& 0.07 & 0.08 &1.8&$1.2 \times 10^{22}$&&1,11\\
&&&& $-76$&8& 0.08 & 0.05 &1.2&$6.9 \times 10^{21}$&&\\
&&&& $-66$&6& 0.06 & 0.04 &1.8&$6.8 \times 10^{21}$&&\\
S51	&	332.666	&$-0.615$& $-49$&$-50$ &28& 0.08 & 0.07 &2.2&$4.1 \times 10^{22}$&&1,12\\
&&&& $-43$&12& 0.08 & 0.08 &1.5&$1.0 \times 10^{22}$&&\\
&&&& $-55$&30& 0.06 & 0.07 &1.7&$2.2 \times 10^{22}$&&\\
S54	&	332.314	&$-0.565$& $-52$ &$-48$&26& 0.07 & 0.06 &2.7&$2.3 \times 10^{22}$&&1\\
&&&& $-44$&11& 0.08 & 0.07 &1.3&$8.8 \times 10^{21}$&&\\
S62	&	331.318	&$-0.356$& $-64$ &$-68$ &27& 0.09 & 0.07 &2.7&$3.4 \times 10^{22}$&&1,13\\
&&&& $-50$&17& 0.08 & 0.09 &3.4&$1.9 \times 10^{22}$&&\\
S64	& 331.005&$-0.148$& $-89$ &$-90$&22& 0.08 & 0.08 &3.5&$4.4\times 10^{22}$&&1\\
&&&& $-102$&6& 0.07 & 0.08 &1.7&$4.1 \times 10^{21}$&&\\
&&&& $-80$&9& 0.07 & 0.07 &1.2&$7.0 \times 10^{21}$&&\\
S66	&	330.784	&	$-0.414$& $-59$ &$-63$&30& 0.11 & 0.12 &2.2&$4.2\times 10^{22}$&&1\\
&&&& $-52$&15&  0.14& 0.08 &2.7&$2.0\times 10^{22}$&&\\
&&&& $-43$&11& 0.14 & 0.09 &2.5&$1.2\times 10^{22}$&&\\
S70	&	329.275	&${+0.112}$& $-77$ &$-77$ &15& 0.07 & 0.09 &3.0&$1.1\times 10^{22}$&&1\\
&&&& $-64$&8& 0.09 & 0.04 &1.8&$8.3\times 10^{21}$&&\\
S71	&	327.986	&	$-0.109$ &$-45$& $-48$&20& 0.08 & 0.07 &2.3&$1.8\times 10^{22}$&&1\\
&&&& $-39$&10& 0.08 & 0.08 &2.0&$1.2 \times 10^{22}$&&\\
S73	&	327.547	& $-0.814$ &$-37$& $-37$ &34&0.07  &0.07  &1.4&$2.6\times 10^{22}$&&1\\
&&&& $-46$&13& 0.04 & 0.06 &1.4&$1.0 \times 10^{22}$&&\\
S74	&	327.527	& $-0.848$&$-37$ &$-37$ &34& 0.07 & 0.07 &1.4&$2.6\times 10^{22}$&&1\\
&&&& $-46$&13& 0.04 & 0.06 &1.4&$1.0 \times 10^{22}$&&\\
    \hline
\end{tabular}}}
\label{tab2}
\end{table*} 
\addtocounter{table}{-1}

\begin{table*}
\tbl{(Continued.)}{
  \centering
   \scalebox{0.95}[0.95]{
  \begin{tabular}{cccccccccccccc}
    \hline
Name & $l$ & $b$& $V_{\rm LSR}^{\rm H\emissiontype{II}}$ & $V_{\rm LSR}^{\rm CO}$&$T_{\rm peak}$&$\sigma_l$&$\sigma_b$ &$\sigma_v$ & $N({\rm H2})_{\rm max}$  & Other & References.\\
     &  [deg] & [deg]&[km s$^{-1}$] &[km s$^{-1}$] &[K]& [deg]& [deg] &[km s$^{-1}$]& [cm$^{-2}$]  & Name&\\
(1) &(2) & (3) & (4) & (5) & (6) & (7) & (8) & (9) & (10) &(11) & (12)\\   
    \hline 
S76	&	326.927	& $-0.021$&$-64$ &$-55$ &19& 0.10 & 0.09 &3.5&$2.8\times 10^{22}$&&1\\
&&&& $-45$&12& 0.09 & 0.10 &2.3&$1.5\times 10^{22}$&&\\
&&&& $-70$&9& 0.10 & 0.10 &2.6&$7.3\times 10^{21}$&&\\
&&&& $-39$&10& 0.09 & 0.08 &1.4&$8.8\times 10^{21}$&&\\
S79	&	326.698	& ${+0.515}$& $-42$ &$-40$ &41& 0.08 & 0.08 &1.8&$4.8\times10^{22}$&&1\\
&&&& $-49$&13& 0.09 & 0.06 &1.8&$1.0\times10^{22}$&&&\\
S91	&	320.868	& $-0.417$ &$-44$&$-44$&17& 0.04 & 0.05 &1.4&$1.8\times10^{22}$&&1\\
&&&& $-61$&20& 0.10 & 0.09 &2.8&$1.9\times10^{22}$&&\\
&&&& $-66$&10& 0.07 & 0.09 &3.5&$1.6\times10^{22}$&&\\
S92	&	320.595	& ${+0.163}$&$-60$&$-60$ &13& 0.09 & 0.07 &1.4&$1.3\times10^{22}$&&1\\
&&&& $-69$&11& 0.12 & 0.12 &2.6&$1.4\times10^{22}$&&\\
S96	&	320.164	&${+0.795}$& $-37$ &$-38$ &32&0.06 & 0.07 &2.6&$3.7\times10^{22}$&&1\\
S97	&	319.886	&	${+0.77}$ &$-38$&$-43$&19& 0.08 & 0.07 &1.9&$1.6\times10^{22}$&&1,14\\
&&&& $-32$&9& 0.08 & 0.04 &0.7&$4.3\times10^{21}$&&\\
S104 &	317.995	 & $-0.750$& $-37$&$-33$&19& 0.05 & 0.05 &1.8&$1.7\times10^{22}$&&1\\
&&&& $-47$&12& 0.10 & 0.07 &2.0&$1.1\times10^{22}$&&\\
S109	&	316.818	&$-0.101$& $-37$&$-39$&27& 0.09 & 0.10 &3.4&$4.5\times 10^{22}$&&1,8,15\\
&&&& $-19$&8& 0.11 & 0.11 &3.0&$9.0\times10^{21}$&&\\
&&&& $-53$&8& 0.12 & 0.13 &2.1&$6.7\times10^{21}$&&\\
S110 &	316.806	&$-0.032$& $-37$& $-39$&27& 0.09 & 0.10 &3.4&$4.5\times 10^{22}$&&1,8,15\\
&&&& $-19$&8& 0.11 & 0.11 &3.0&$9.0\times10^{21}$&&\\
&&&& $-53$&8& 0.12 & 0.13 &2.1&$6.7\times10^{21}$&&\\
S111&	316.772	&$-0.075$& $-37$& $-39$&27& 0.09 & 0.10 &3.4&$4.5\times 10^{22}$& & 1,8,15\\
&&&& $-19$&8& 0.11 & 0.11 &3.0&$9.0\times10^{21}$&&\\
&&&& $-53$&8& 0.12 & 0.13 &2.1&$6.7\times10^{21}$&&\\
S116&314.237 &	${+0.477}$ &$-60$&$-61$ &22& 0.09 & 0.11 &2.8&$3.0\times 10^{22}$&&1,16\\
&&&& $-46$&12& 0.12 & 0.12 &2.2&$1.6\times 10^{22}$&&\\
S123	&	312.976	&	$-0.436$ & $-47$&$-44$&14& 0.09 & 0.08 &2.9&$1.6\times10^{22}$&&1\\
&&&& $-34$&16& 0.09 & 0.08 &1.8&$1.4\times 10^{22}$&&\\
S133	&	311.487	&${+0.398}$ &$-62$ &$-59$ &20& 0.08 & 0.08 &4.6&$2.9 \times 10^{22}$&&1\\
&&&& $-35$&10&0.07  & 0.08 &2.3&$1.3 \times 10^{22}$&&\\
S137	&	310.981	& ${+0.41}$ &$-53$ &$-54$&17& 0.07 & 0.06 &2.0 &$1.4 \times 10^{22}$&RCW 82&1,17\\
&&&& $-46$&22& 0.08 & 0.07 &1.5&$1.4\times10^{22}$&&\\
&&&& $-39$&9& 0.07 & 0.08 &2.4&$7.4 \times 10^{21}$&&\\
S141	&	309.552	&	$-0.721$& $-43$&$-53$&16& 0.07 & 0.07 &1.6&$1.8 \times 10^{22}$&&1\\
&&&& $-42$ &18& 0.06 & 0.07 &2.1&$2.4 \times 10^{22}$&&\\
&&&& $-26$&11& 0.09 & 0.08 &1.3&$7.5 \times 10^{21}$&&\\
S143	&	309.05	&${+0.157}$&$-47$ &$-43$&19& 0.13 &  0.14&3.2&$2.3\times10^{22}$&&1\\
&&&& $-50$&26& 0.14 & 0.12 &2.5&$2.3\times10^{22}$&&\\
S145	&	308.701	&${+0.641}$& $-46$&$-47$&24& 0.11 & 0.15 &1.9&$2.0\times10^{22}$& RCW 79&1 18,19,20,21\\
&&&& $-40$&10& 0.16 & 0.17 &1.2&$6.7\times10^{21}$&&\\
&&&& $-57$&7& 0.10 & 0.18 &1.1&$4.6\times10^{21}$&&&\\
S150	&	305.534	&	${+0.357}$& $-38$&$-38$&21& 0.05 & 0.05 &3.2&$2.3\times10^{22}$&&1\\
&&&& $-47$&10& 0.04 & 0.06 &1.3&$6.6\times10^{21}$&&&\\
S156	&	305.261	&	${+0.215}$& $-39$ &$-39$&31& 0.06 & 0.05&4.0&$5.2\times10^{22}$&&1,22\\
S163	&	303.893	& $-0.704$& $+30$ &$+30$ &12& 0.05 & 0.05 &2.2&$1.6\times10^{22}$&&1\\
S181	&	298.218	& $-0.315$& $+31$ &$+34$ &28& 0.05 & 0.07 &2.8&$3.4 \times 10^{22}$&&1\\
&&&& $+23$&8& 0.06 & 0.08 &1.4&$4.8\times10^{21}$&&&\\
S186	&	295.177	& $-0.661$ &$+38$&$+35$ &17& 0.06 & 0.08 &2.5&$1.4\times10^{22}$&&23\\
&&&& $+27$&11& 0.09 & 0.07 &1.1&$9.3\times10^{21}$&&\\
&&&& $+21$&7& 0.07 & 0.09 &1.8&$3.7\times10^{21}$&&\\
&&&& $+16$&10& 0.06 & 0.06 &0.8&$6.0\times10^{21}$&&\\
    \hline
\end{tabular}}}
\label{tab2}
Columns: (1) Region name in \citet{2006ApJ...649..759C},  (2) Galactic longitude in the center position of the infrared bubble, (3) Galactic latitude in the center position of the infrared bubble, (4) Radial velocity of a H\emissiontype{II} region obtained by previous studies. (5) Radial velocity of a molecular cloud at the peak position. (6) Peak temperature of $^{12}$CO $J=$1-0. (7) Intensity-weighted standard deviation of the Galactic longitude (8) Intensity-weighted standard deviation of the Galactic latitude (9) Intensity-weighted standard deviation of the radial velocity (10) Peak H$_2$ column density assuming the CO-to-H$_2$ conversion factor.  (11) Other region names (12) References of the radial velocity of H\emissiontype{II} regions and other CO-line observations. 
[1] \citet{2014MNRAS.438..426H}
[2] \citet{2007A&A...472..835Z}
[3] \citet{2015ApJ...800..101A}
[4] \citet{2015ApJ...806....7T}
[5] \citet{2017A&A...600A..93F}
[6] \citet{1999MNRAS.308..683C}
[7] \citet{2000A&A...361.1079Z}
[8] \citet{2018A&A...617A..67S}
[9] \citet{2017ApJ...840..111T}
[10] \citet{2010ApJ...719.1104R}
[11] \citet{2021PASJ...73S.338K}
[12]\citet{2012A&A...544A..11Z}
[13]\citet{2018ApJ...862...10M}
[14]\citet{2015A&A...582A...1D}
[15]\citet{2022ApJ...926....4M}
[16] \citet{2018PASJ...70S..46F}
[17] \citet{2009A&A...494..987P}
[18] \citet{2006A&A...446..171Z}
[19] \citet{2018PASJ...70S..45O}
[20] \citet{2017A&A...602A..95L}
[21] \citet{2023A&A...679L...5B}
[22] \citet{2022ApJ...928...83Y}
[23] \citet{1987A&A...171..261C}

\end{table*}

\begin{table*}
\tbl{Kinematic distances, diameters, size parameters, total molecular mass, and virial mass of molecular clouds toward the infrared bubbles in the Southern Galactic plane.}{
\centering
\scalebox{0.95}[0.95]{
\begin{tabular}{cccccccccccccc}
\hline
Name & $r$ & $V_{\rm LSR}^{\rm CO}$ &$d_{\rm near}$ & $d_{\rm far}$& $\delta d $ & $D_{\rm near}$ & $D_{\rm far}$& $S_{\rm near}$ &$S_{\rm far}$&$M^{\rm near}_{\rm cloud}$& $M^{\rm far}_{\rm cloud}$&$M^{\rm near}_{\rm vir}$& $M^{\rm far}_{\rm vir}$ \\
      &[$\timeform{'}$] & [km s$^{-1}$] &[kpc] & [kpc] & [kpc] & [pc] & [pc] & [pc] & [pc] &[$M_{\odot}$] &[$M_{\odot}$]&[$M_{\odot}$] &[$M_{\odot}$]\\
(1) &(2) & (3) & (4) & (5) & (6) & (7) & (8) & (9)& (10) & (11) & (12) & (13) & (14)\\   
\hline
S1 	& 4.33& $-26$& 3.4& 12.2& 0.5 & 8.5 & 30.8 &3.8&13.9& $2.8 \times 10^4$ &$3.7 \times 10^5$&$2.7 \times 10^4$ &$9.7 \times 10^4$\\
&&$-17$ &2.4 & 13.2& 0.6 & 6.1& 33.2& 3.9 & 21.0 &$1.2 \times 10^4$&$3.6 \times 10^5$&$2.1 \times 10^4$&$1.2 \times 10^5$\\
S7 & 4.45 &$-7$& 0.9 & 14.6 & 0.9 & 2.3& 37.8 &1.4&23.0&$8.1 \times 10^3$ & $2.2 \times 10^6$&$1.2 \times 10^4$ & $2.0 \times 10^5$\\
&& $-28$& 3.2 & 12.3 & 0.4 & 8.3 & 31.9& 2.4& 9.3&$7.9 \times 10^3$&$1.2 \times 10^5$&$5.9 \times 10^3$&$2.3 \times 10^4$\\
S8 	&1.64 &$-90$& 6.0 & 9.5 & 0.1 & 5.7&9.1 &6.1&9.6& $1.9 \times 10^4$&$4.9 \times 10^5$& $5.4 \times 10^4$&$8.6 \times 10^4$\\
&& $-83$& 5.8&9.7 & 0.1 & 5.5 & 9.3 &7.0&11.7&$2.5 \times 10^4$ &$7.1 \times 10^4$&$1.2 \times 10^5$ &$2.1 \times 10^5$\\
S11	& 2.33&$-18$ & 1.9 & 13.5 & 0.4 & 2.5 & 18.3 &2.6&18.6& $3.7 \times 10^4$&$1.9 \times 10^6$& $4.6 \times 10^4$&$3.3 \times 10^5$\\
& & $-40$& 3.7 &  11.7 & 0.4 & 5.0 & 15.8 &4.8&15.2& $4.3\times 10^4$ & $4.3 \times 10^5$& $6.3\times 10^4$ & $2.0 \times 10^5$\\ 
S13	& 8.76&$-21$ & 2.1 & 13.2 & 0.4  & 10.8 & 67.2 &4.9&30.6& $6.6 \times 10^4$&$2.5 \times 10^6$& $8.6 \times 10^4$&$5.3 \times 10^5$\\
&& $-10$& 1.1& 14.3& 0.6 &5.4 & 72.6&1.5&20.4& $7.6 \times 10^2$&$1.4 \times 10^5$& $2.0 \times 10^3$&$2.7 \times 10^4$\\
&& $-30$& 2.8& 12.5& 0.3 & 14.4 & 63.6&4.9&21.5& $1.5 \times 10^4$&$2.9 \times 10^5$& $2.5 \times 10^4$&$1.1 \times 10^5$\\
S14	& 4.88& $-33$& 3.1& 12.2& 0.3 & 8.7 & 34.7 & 5.8 &23.1&$2.3 \times 10^4$ &$3.6 \times 10^5$&$2.1 \times 10^4$ &$8.5 \times 10^4$\\
&& $-24$& 2.4 & 12.9 & 0.4 & 6.7 & 36.7 &4.6&24.9&$4.5 \times 10^4$  &$1.3 \times 10^6$ &$5.5 \times 10^4$  &$3.0 \times 10^5$ \\
&& $-17$& 1.7& 13.5 & 0.5 & 4.9 & 38.5 &3.1&23.8&$1.3 \times 10^4$ &$7.9 \times 10^5$&$2.5 \times 10^4$ &$1.9 \times 10^5$ \\
S15	& 2.18&$-27$ & 2.5& 12.7 & 0.4 & 3.2& 16.1 &3.6&18.3& $7.0 \times 10^4$&$1.8 \times 10^6$& $1.2 \times 10^5$&$6.0 \times 10^5$\\
&& $-4$& 0.3 & 14.9 & 0.8 & 0.4 & 18.9 &0.4&20.7&$9.6 \times 10^1$&$2.3 \times 10^5$& $1.7 \times 10^3$&$8.4\times 10^4$\\
&& $-36$& 3.2 & 12.0 & 0.3 & 4.0 & 15.3 &3.2&12.2&$8.6 \times10^3$&$1.2 \times 10^5$&$1.0 \times10^4$&$3.9 \times 10^4$\\
S17	&1.81&$-30$ & 2.7 & 12.4 & 0.5 & 2.9 & 13.1 &3.8&17.1&$3.5 \times 10^4$ &$7.2 \times 10^5$&$5.8 \times 10^4$ &$2.6 \times 10^5$\\
&& $-36$& 3.1 & 12.1 & 0.3 & 3.3& 12.7&4.0&15.6&$7.7 \times 10^3$ &$1.2 \times 10^5$&$1.5 \times 10^4$ &$5.8 \times 10^4$\\
&& $-19$& 1.8 & 13.4 & 0.4 & 1.8& 14.1&2.2&16.7&$2.0 \times 10^3$ &$1.2 \times 10^5$&$7.9 \times 10^3$ &$6.0 \times 10^4$\\
S18	& 1.34 &$-71$& 4.8& 10.3 & 0.2 & 3.7& 8.0 & 5.9 & 12.6 & $4.3 \times 10^4$&$2.0 \times 10^5$& $3.9 \times 10^4$&$8.4 \times 10^4$\\
&& $-79$& 5.1& 10.0 & 0.2 & 4.0 & 7.8& 7.3 & 14.3 & $9.2 \times 10^4$&$3.5 \times 10^5$& $5.2 \times 10^4$&$1.0 \times 10^5$\\
&& $-84$& 5.3& 9.8 & 0.2 & 4.1 & 7.6& 7.5 & 14.0 & $2.7 \times 10^4$&$9.4 \times 10^4$& $2.1 \times 10^4$&$4.0 \times 10^4$\\
S20	& 1.53 &$-71$& 4.8& 10.3 & 0.2 & 4.3& 9.1 & 5.9 & 12.6 & $4.3 \times 10^4$&$2.0 \times 10^5$& $3.9 \times 10^4$&$8.4 \times 10^4$\\
&& $-79$& 5.1& 10.0 & 0.2 & 4.5 & 8.9& 7.3 & 14.3& $9.2 \times 10^4$&$3.5 \times 10^5$& $5.2 \times 10^4$&$1.0 \times 10^5$\\
&& $-84$& 5.3& 9.8 & 0.2 & 4.7 & 8.7& 7.5 & 14.0 &  $2.7 \times 10^4$&$9.4 \times 10^4$& $2.1 \times 10^4$&$4.0 \times 10^4$\\
S23	& 2.06&$-45$ & 3.4 & 11.6 & 0.2 & 4.1 & 13.9 &5.0&17.2&$1.3 \times 10^5$ & $1.6 \times 10^6$&$1.5 \times 10^5$ & $5.3 \times 10^5$\\
&& $-30$& 2.4 & 12.6 & 0.3 & 2.9 & 15.1 &3.4&17.7& $3.2 \times 10^4$&$8.5 \times 10^5$& $6.1 \times 10^4$&$3.2 \times 10^5$\\
S27	& 1.70&$-53$ & 3.8& 11.1 & 0.3 & 3.7 & 11.0 &4.4&12.9& $1.2 \times 10^5$&$1.0 \times 10^6$& $1.2 \times 10^5$&$3.6 \times 10^5$\\
&& $-39$& 3.0 & 11.9 & 0.2 & 2.9 & 11.8& 3.9& 15.5&$1.4 \times 10^4$ &$2.2 \times 10^5$&$4.3 \times 10^4$ &$1.7 \times 10^5$\\
S29	& 2.52&$-66$ & 4.3& 10.5& 0.2 & 6.3 & 15.4 &5.9&14.5& $2.1 \times 10^5$ &$1.2 \times 10^6$& $1.3 \times 10^5$ &$3.3 \times 10^5$ \\
&& $-71$& 4.5 & 10.3 & 0.2 & 6.6 & 15.0 &4.6&10.5&$1.6 \times 10^4$ &$8.5 \times 10^4$&$3.1 \times 10^4$ &$7.1 \times 10^4$\\
&& $-52$& 3.6 & 11.2 & 0.2 & 5.3 & 16.4 &5.6&17.3&$7.6 \times 10^3$ &$7.3 \times 10^4$&$1.9 \times 10^4$ &$6.0 \times 10^4$\\
S36	& 3.00& $-41$& 2.9 & 11.8 & 0.3 & 5.1 & 20.6 &4.3&17.4& $2.7 \times 10^4$&$4.5 \times 10^5$& $1.1 \times 10^5$&$4.4 \times 10^5$\\
&& $-23$& 1.7 & 13.0 & 0.3 & 3.0 & 22.7&1.6 & 12.3&$3.1 \times 10^2$ &$1.8 \times 10^4$&$1.1 \times 10^4$ &$8.4 \times 10^4$\\
&& $-53$& 3.6 & 11.0 & 0.4 & 6.4 & 19.3&3.5 &10.7&$1.3 \times 10^4$ &$1.2 \times 10^5$&$8.5 \times 10^4$ &$2.6 \times 10^5$\\
S37	& 1.30&$-30$& 2.3 & 12.4 & 0.4 & 1.7 & 9.4 &3.2&17.8&$2.3 \times 10^4$ &$6.9 \times 10^5$&$1.2 \times 10^5$ &$6.4 \times 10^5$\\
&& $-54$& 3.7 & 11.0 & 0.2 & 2.8 & 8.3 &4.7&14.3&$2.6 \times 10^4$ &$2.4 \times 10^5$&$1.5 \times 10^5$ &$4.4 \times 10^5$\\
S41	& 3.35&$-84$& 5.0 & 9.5 & 0.3 & 9.7& 18.6&5.4& 10.2& $5.9 \times 10^4$&$2.2 \times 10^5$& $7.6 \times 10^4$&$1.5 \times 10^5$\\
&& $-79$& 4.7 & 9.8 & 0.3 & 9.2 & 19.1 &5.8&12.1& $2.3 \times 10^4$&$9.7 \times 10^4$& $6.7 \times 10^4$&$1.4 \times 10^5$\\
S44	& 2.55&$-80$& 4.8&9.5 & 0.3 & 7.1& 14.1 &6.2&12.2&$4.2 \times 10^4$ &$1.6 \times 10^5$&$4.2 \times 10^4$ &$8.3 \times 10^4$\\
&& $-76$& 4.5 & 9.8 & 0.3 & 6.7 & 14.5 &5.0&10.8& $1.6 \times 10^4$&$7.4 \times 10^4$& $1.5 \times 10^4$&$3.2 \times 10^4$\\
&& $-66$& 4.1 & 10.2 & 0.3 & 6.1 & 15.1 &3.4&8.4& $3.9 \times 10^3$&$2.4 \times 10^4$& $2.1 \times 10^4$&$5.2 \times 10^4$\\
S51	& 1.68&$-50$& 3.1& 10.9& 0.3& 3.1 & 10.7&4.1&14.3& $7.2 \times 10^4$ &$8.9 \times 10^5$& $4.0 \times 10^4$ &$1.4 \times 10^5$\\
&& $-43$& 2.7 & 11.4 & 0.3 & 2.6& 11.1 & 3.7&15.7&$1.4 \times 10^4$ &$2.6 \times 10^5$&$1.7 \times 10^4$ &$7.3 \times 10^4$\\
&& $-55$& 3.4 & 10.7 & 0.3 & 3.3& 10.4 & 4.0&12.6&$3.1 \times 10^4$ &$3.1 \times 10^5$&$2.3 \times 10^4$ &$7.0 \times 10^4$\\
S54	& 1.08&$-48$& 3.0 & 11.0 & 0.3 & 1.9 & 6.9 &3.2&11.8& $2.4 \times 10^4$&$3.3\times 10^5$& $4.8\times 10^4$&$1.8\times 10^5$\\
&& $-44$& 2.8 & 11.2 & 0.3 & 1.8 & 7.0& 3.8& 14.9 &$1.6\times 10^4$ & $2.5 \times 10^5$&$1.3\times 10^4$ & $4.9 \times 10^4$\\
S62	& 1.90&$-68$ & 4.1& 9.8& 0.3 & 4.5& 10.8&5.6&13.2&$1.3 \times 10^5$ &$7.5 \times 10^5$&$8.3 \times 10^4$ &$2.0 \times 10^5$\\
&& $-50$& 3.0& 10.9 & 0.3 & 3.4 & 12.0&4.5&16.1& $3.2 \times 10^4$&$4.0 \times 10^5$&$1.1 \times 10^5$ &$3.8 \times 10^5$\\
S64	& 3.03&$-90$& 5.4&8.4 & 0.4 & 9.5 & 14.9&7.6&11.9& $3.0 \times 10^5$&$7.2 \times 10^5$& $1.9 \times 10^5$&$2.9 \times 10^5$\\
&& $-102$& 6.3& 7.6 & 0.4 &11.1 &13.3 &8.2 & 9.8&$4.1 \times 10^3$ &$6.0 \times 10^3$&$4.6 \times 10^4$ &$5.6 \times 10^4$\\
&& $-80$& 4.6& 9.2 & 0.2 &8.2 &16.2 &5.4 & 10.8&$7.8 \times 10^3$ &$3.1 \times 10^4$&$1.5 \times 10^4$ &$3.0 \times 10^4$\\
S66	& 6.31&$-63$& 3.8&10.0 & 0.3 & 14.1& 36.6&7.6 &19.8&$2.0 \times 10^5$&$1.4 \times 10^6$&$7.5 \times 10^4$&$2.0 \times 10^5$\\
&& $-52$& 3.2& 10.6& 0.3 & 11.8& 38.9&5.8&18.9& $6.9\times 10^4$&$7.4 \times 10^5$& $8.2\times 10^4$&$2.7 \times 10^5$\\
&& $-43$& 2.6& 11.2& 0.2 & 9.6& 41.2&4.9&21.1& $2.0\times 10^4$&$3.7 \times 10^5$& $6.0\times 10^4$&$2.6 \times 10^5$\\
S70	& 1.13&$-77$ & 4.5& 9.1 & 0.2 & 3.0 & 6.0 & 6.1&12.2&$1.3\times 10^4$&$5.1\times 10^4$&$1.1\times 10^5$&$2.2\times 10^5$\\
 & &$-64$ & 3.8& 9.8 & 0.3 & 2.5 & 6.4 & 3.9 & 10.1&$8.8\times 10^3$&$5.8\times 10^4$&$2.6\times 10^4$&$6.6\times 10^4$\\
S71	& 1.07&$-48$&2.9 &10.5 &0.3  &1.8 &6.6 &3.7 &13.5 &$2.0 \times 10^4$&$2.6 \times 10^5$&$4.0 \times 10^4$&$1.5 \times 10^5$\\
&& $-39$&2.4 &11.0 &0.3  &1.5 & 6.9& 3.4&15.7 &$1.8 \times 10^4$ & $3.9 \times 10^5$&$2.7 \times 10^4$ & $1.2 \times 10^5$\\
S73	& 3.88 & $-37$ &2.3 &11.1 &0.3  & 5.1& 25.1& 2.9&14.4 &$1.1 \times 10^4$&$2.7 \times 10^5$&$1.2 \times 10^4$&$6.0 \times 10^4$\\
&& $-46$& 2.8& 10.6& 0.3 &6.2 &23.9 &2.3 &8.7 &$2.4 \times 10^3$ &$3.5 \times 10^4$&$9.5 \times 10^3$ &$3.6 \times 10^4$\\
S74	& 1.29 &$-37$ & 2.3&11.1 & 0.3 &1.7 &8.3 & 2.9&14.4 &$1.1 \times 10^4$&$2.7 \times 10^5$&$1.2 \times 10^4$&$6.0 \times 10^4$\\
&& $-46$&2.8 &10.6 & 0.3 & 2.1&8.0 &2.3 &8.7 &$2.4 \times 10^3$ & $3.5 \times 10^4$ & $9.5 \times 10^3$ & $3.6 \times 10^4$\\
\hline
\end{tabular}}}
\label{tab3}
\end{table*} 
\addtocounter{table}{-1}

\begin{table*}
\tbl{(Continued.)}{
  \centering
   \scalebox{0.95}[0.95]{
  \begin{tabular}{cccccccccccccc}
    \hline
Name  &$r$ & $V_{\rm LSR}^{\rm CO}$ &$d_{\rm near}$ & $d_{\rm far}$ &$\delta d$ & $D_{\rm near}$ & $D_{\rm far}$& $S_{\rm near}$ & $S_{\rm far}$ &$M^{\rm near}_{\rm cloud}$& $M^{\rm far}_{\rm cloud}$ &$M^{\rm near}_{\rm vir}$& $M^{\rm far}_{\rm vir}$ \\
      & [$\timeform{'}$] & [km s$^{-1}$] & [kpc] &[kpc] & [kpc] & [pc] & [pc] & [pc] & [pc] &[$M_{\odot}$] &[$M_{\odot}$] &[$M_{\odot}$] &[$M_{\odot}$]\\
(1) &(2) & (3) & (4) & (5) & (6) & (7) & (8) & (9)& (10) & (11) & (12) & (13) & (14)\\    
    \hline 
S76	& 5.14 &$-55$ & 3.3&10.0 & 0.4 & 9.9& 29.8&5.2 &15.7 & $9.8 \times 10^4$ & $8.8 \times 10^5$& $1.3 \times 10^5$ & $3.8 \times 10^5$\\
&& $-45$& 2.6 & 10.7 & 0.2 & 7.8 & 31.8& 4.4&17.7 &$4.5 \times 10^4$&$7.4 \times 10^5$&$4.7 \times 10^4$&$1.9 \times 10^5$\\
&& $-70$& 4.1& 9.2& 0.2 & 12.1& 27.6& 7.3&16.6 &$2.0 \times 10^4$ & $1.0 \times 10^5$&$9.7 \times 10^4$ & $2.2 \times 10^5$\\
&& $-39$&2.3 &11.0 & 0.2 & 6.8& 32.9&3.5 &16.9&$1.1 \times 10^4$ & $2.6 \times 10^5$&$1.3 \times 10^4$ & $6.3 \times 10^4$\\
S79	& 2.79&$-40$ &2.5&10.8 &0.3 & 4.0 & 17.5& 3.3&14.6 &$5.6 \times 10^4$ & $1.1 \times 10^6$&$2.2 \times 10^4$ & $9.6 \times 10^4$\\
&& $-49$& 3.0&10.3 &0.3 & 4.8 & 16.6& 3.8&13.1 &$7.6 \times 10^3$ & $9.0 \times 10^4$&$2.4 \times 10^4$ & $8.2 \times 10^4$\\
S91	& 2.56 &$-44$& 2.6& 9.7& 0.4 & 3.8 & 14.5& 2.1& 7.7 & $5.8 \times 10^3$ & $8.2 \times 10^4$& $7.7 \times 10^3$ & $2.9 \times 10^4$\\
&& $-61$& 3.7& 8.6& 0.2 & 5.5 & 12.8& 6.0& 13.9 & $3.2 \times 10^4$ & $1.7 \times 10^5$& $9.5 \times 10^4$ & $2.2 \times 10^5$\\
&& $-66$& 4.0& 8.3& 0.3 & 5.9 & 12.4& 5.6& 11.7 & $5.7 \times 10^4$ & $2.5 \times 10^5$& $1.3 \times 10^5$ & $2.8 \times 10^5$\\
S92	& 5.49 &$-60$ &3.8 & 8.5& 0.5 & 12.0& 27.1&5.4 &12.2 & $1.5 \times 10^4$ & $7.5 \times 10^4$& $2.2 \times 10^4$ & $5.0 \times 10^4$\\
&& $-69$& 4.3& 7.9& 0.3 & 13.8 & 25.3& 9.2& 16.9 & $9.8 \times 10^4$ & $3.3 \times 10^5$& $1.3 \times 10^5$ & $2.3 \times 10^5$\\
S96	& 1.40 & $-38$ & 2.2&10.0 & 0.4 &1.8 & 8.1&2.6 &11.6 &$1.8 \times 10^4$ & $3.7 \times 10^5$&$3.4 \times 10^4$ & $1.5 \times 10^5$\\
S97	& 1.28 &$-43$&2.4 &9.7 & 0.2 &1.8 &7.2 &3.2 &12.8&$1.9 \times 10^4$ & $3.1 \times 10^5$&$2.3 \times 10^4$ & $9.1 \times 10^4$\\
&& $-32$& 1.8& 10.3& 0.4 & 1.3 & 7.7& 1.8& 10.6 & $1.0 \times 10^3$ & $3.4 \times 10^4$& $1.6 \times 10^3$ & $9.4 \times 10^3$\\
S104 & 2.01& $-33$& 1.9&9.9 & 0.4 &2.2 &11.6 & 1.6&8.3 & $1.5 \times 10^3$ & $4.3 \times 10^4$& $1.1 \times 10^4$ & $5.7 \times 10^4$\\
&& $-47$&2.8 &9.0 & 0.2 &3.3 &10.5 & 4.0&12.8 & $4.7 \times 10^3$ & $4.9 \times 10^4$& $3.2 \times 10^4$ & $1.0 \times 10^5$\\
S109 & 1.59&$-39$& 2.4& 9.1& 0.5 &2.2 & 8.5& 4.1&15.5&$8.2 \times 10^4$&$1.2 \times 10^6$&$9.6 \times 10^4$&$3.7 \times 10^5$\\
&& $-19$&0.8 &10.7 & 0.2 & 0.8& 9.9& 1.6&21.0&$8.0 \times 10^2$&$1.3 \times 10^5$&$3.0 \times 10^4$&$3.8 \times 10^5$\\
&& $-53$& 3.2& 8.4& 0.3 & 2.9&7.8 &6.8 &18.0 &$2.7 \times 10^3$&$1.9 \times 10^4$&$6.1 \times 10^4$&$1.6 \times 10^5$\\
S110 & 1.59&$-39$& 2.4& 9.1& 0.5 &2.2 & 8.5& 4.1&15.5&$8.3 \times 10^4$&$1.2 \times 10^6$&$9.6 \times 10^4$&$3.7 \times 10^5$\\
&& $-19$&0.8 &10.7 & 0.2 & 0.8& 9.9& 1.6&21.0&$8.0 \times 10^2$&$1.3 \times 10^5$&$3.0 \times 10^4$&$3.8 \times 10^5$\\
&& $-53$& 3.2& 8.4& 0.3 & 2.9&7.8 &6.8 &18.0 &$2.7 \times 10^3$&$1.9 \times 10^4$&$6.1 \times 10^4$&$1.6 \times 10^5$\\
S111 &1.52 &$-39$& 2.4& 9.1& 0.5 &2.1 & 8.1& 4.1&15.5&$8.3 \times 10^4$&$1.2 \times 10^6$&$9.6 \times 10^4$&$3.7 \times 10^5$\\
&& $-19$&0.8 &10.7 & 0.2 & 0.7& 9.5& 1.6&21.0&$8.0 \times 10^2$&$1.3 \times 10^5$&$3.0 \times 10^4$&$3.8 \times 10^5$\\
&& $-53$& 3.2& 8.4& 0.3 & 2.8&7.4 &6.8 &18.0 &$2.7 \times 10^3$&$1.9 \times 10^4$&$6.1 \times 10^4$&$1.6 \times 10^5$\\
S116 & 3.85 &$-61$ &4.4 & 6.6& 0.5 & 9.9& 14.9&7.5 &11.3 &$1.5 \times 10^5$&$3.3 \times 10^5$&$1.2 \times 10^5$&$1.8 \times 10^5$\\
&& $-46$&2.9 &8.2 & 0.3 &6.5 &18.3 & 6.0&17.1 &$3.1 \times 10^4$&$2.5 \times 10^5$&$5.7 \times 10^4$&$1.6 \times 10^5$\\
S123 & 2.32 &$-44$& 2.7& 8.1& 0.3 & 3.7&10.9 &4.0 &11.9&$2.8 \times 10^4$&$2.5 \times 10^5$&$6.7 \times 10^4$&$2.0 \times 10^5$\\
&& $-34$& 1.9& 8.9& 0.3 & 2.5& 12.0&2.8 &13.1&$8.8 \times 10^3$&$2.0 \times 10^5$&$1.9 \times 10^4$&$8.9 \times 10^4$\\
S133 & 2.03 &$-59^*$ & 5.2& --- & --- & 6.2& ---& 7.5&---&$2.7 \times 10^5$&---&$3.3 \times 10^5$&---\\\
&& $-35$&2.0 &8.5 & 0.3 & 2.3&10.1 &2.7 &11.6&$7.0 \times 10^3$&$1.3 \times 10^5$&$2.8 \times 10^4$&$1.2 \times 10^5$\\
S137 & 3.43 &$-54$& 3.8 & 6.6 & 0.5 & 7.6  &13.1&4.5& 7.8 &$2.6 \times 10^4$ & $7.6 \times 10^4$&$3.8 \times 10^4$ & $6.6 \times 10^4$\\
&& $-46$&3.0 & 7.4& 0.3 &5.9 &14.8 & 3.8& 9.5 &$2.0 \times 10^4$ & $1.3 \times 10^5$&$1.8 \times 10^4$ & $4.5 \times 10^4$\\
&& $-39$&2.3 & 8.1& 0.3 &4.6 &16.1 & 3.0& 10.5 &$5.1 \times 10^3$ & $6.3 \times 10^4$&$3.3 \times 10^4$ & $1.2 \times 10^5$\\
S141 & 1.46 &$-53$ & 4.4& 5.6& 0.8 &3.8 & 4.8& 5.3& 6.8&$1.9 \times 10^4$ & $3.1 \times 10^4$&$2.8 \times 10^4$ & $3.5 \times 10^4$\\
&& $-42$&2.8 & 7.3& 0.4 & 2.4& 6.2&3.2 &8.4&$2.2 \times 10^4$ & $1.5 \times 10^5$&$2.9 \times 10^4$ & $7.7 \times 10^4$\\
&& $-26$&1.5 & 8.6& 0.5 & 1.3& 7.3&2.3 &12.7&$3.7 \times 10^3$ & $1.2 \times 10^5$&$7.4 \times 10^3$ & $4.2 \times 10^4$\\
S143 & 5.34&$-43$&2.9 &7.1 & 0.4 & 8.9&22.1 & 6.5&16.4&$1.2 \times 10^5$ & $7.4 \times 10^5$&$1.4 \times 10^5$ & $3.4 \times 10^4$\\
&& $-50$& 3.8& 6.1& 0.6 &11.9 & 19.1& 8.4&13.4&$6.9 \times 10^4$ & $1.8 \times 10^5$&$1.0 \times 10^5$ & $1.7 \times 10^5$\\
S145 & 6.36&$-47$& 3.4& 6.6& 0.5 &12.4 &24.2 & 7.6 &14.9 &$4.0 \times 10^4$ & $1.5 \times 10^5$&$5.5 \times 10^4$ & $1.1 \times 10^5$\\
&& $-40$& 2.5& 7.4& 0.3 &9.3 &27.3 &7.3 &21.3&$9.7 \times 10^3$ & $8.3 \times 10^4$&$2.2 \times 10^4$ & $6.4 \times 10^4$\\
&& $-57^*$&5.0 & ---&--- &18.3 & --- & 11.8&--- &$7.9 \times 10^3$ & ---&$2.9 \times 10^4$ & --- \\
S150 & 1.10 &$-38$& 2.5& 6.7& 0.4 &1.6 & 4.3&2.3 &6.2&$1.9 \times 10^4$ & $1.3 \times 10^5$&$4.7 \times 10^4$ & $1.3 \times 10^5$\\
&& $-47^*$&4.6 & ---&--- &2.9 & --- & 3.9& ---&$4.5\times 10^3$ & ---&$1.3\times 10^4$ & --- \\
S156 & 1.93 &$-39$& 3.1& 6.1 & 0.9&3.4 &6.8 &3.1&6.3&$4.5 \times 10^4$ & $1.8 \times 10^5$&$1.0 \times 10^5$ & $2.0 \times 10^5$\\
S163 & 2.94&$+30$ &---&11.8&0.5& ---& 20.1&---  &10.3 & ---& $6.8 \times 10^4$& ---& $1.0 \times 10^5$\\
S181 & 1.79&$+34$ &--- &11.0 & 0.3 & ---& 11.5&--- &10.7&--- & $9.2 \times 10^4$&--- & $1.7 \times 10^5$\\
&& $+23$& ---&9.9 & 0.6& ---& 10.3 &--- &12.1 &--- & $1.2 \times 10^4$&--- & $4.6 \times 10^4$\\
S186 & 5.43 &$+35$ &--- &10.4  & 0.3&--- &33.0 &---&13.4&--- & $3.3 \times 10^4$&--- & $1.7 \times 10^5$\\
&& $+27$& ---& 9.6& 0.6 &--- &30.3 & ---&13.1&--- & $2.2 \times 10^4$&--- & $3.1 \times 10^4$\\
&& $+21$& ---& 9.3& 0.3 &--- &29.4 & ---&12.8&--- & $6.9 \times 10^3$&--- & $8.7 \times 10^4$\\
&& $+16$& ---& 8.7& 0.6 &--- &27.5 & ---&8.5&--- & $1.2 \times 10^4$&---&$1.2 \times 10^4$\\
\hline
\end{tabular}}}
\label{tab3}
Columns: (1) Region name in \citet{2006ApJ...649..759C}, (2) The radii of the infrared bubbles obtained by \citet{2019PASJ...71....6H} (3) Radial velocity of a molecular cloud, (4) Near solution of the kinematic distance for the CO radial velocity, (5) Far solution of the kinematic distance for the CO radial velocity. (6) The distance error (7) A linear diameter of the infrared bubble for the near solution of the kinematic distance. (8) A linear diameter of the infrared bubble for the far solution of kinematic distance. (9) A size parameter for the near solution of the kinematic distance. (10) A size parameter for the far solution of the kinematic distance. (11) Molecular mass for the near solution of the kinematic distance. (12) Molecular mass for the far solution of kinematic distance.  (13) Virial mass for the near solution of the kinematic distance. (14) Virial mass for the far solution of kinematic distance. $^*$ They correspond to the terminal velocity of $R^2 - R_0^2 \sin l=0$.
\end{table*} 

\section*{Supplementary data}
Supplementary figures E1 -- E43 present the catalog of molecular clouds and position-velocity diagrams possibly associated with infrared bubbles, except for the S1 bubble.

\section*{Acknowledgements}
{We are grateful to the anonymous referee for carefully reading our manuscript and giving us thoughtful suggestions, which greatly improved this paper.}
We utilized the Python software package for astronomy \citep{2013A&A...558A..33A,2018AJ....156..123A},
NumPy \citep{2011CSE....13b..22V}, Matplotlib \citep{2007CSE.....9...90H},IPython \citep{2007CSE.....9c..21P}, and APLpy \citep{2012ascl.soft08017R}. 

The authors thank Dr. Michael Burton of the University of New South Wales and the Armagh Observatory and Planetarium for the archival CO survey data with Mopra. 
We also thank Dr. Graeme Wong for kindly supporting remote observations from Nagoya University.
The Mopra radio telescope is part of the Australia Telescope National Facility, which is funded by the Australian Government for operation as a National Facility managed by CSIRO. The University of New South Wales Digital Filter Bank used for the observations with the Mopra Telescope was provided with support from the Australian Research Council.


{}  


\end{document}